\documentstyle[amssymb,twocolumn,aps]{revtex}

\begin{document}
\draft
\title{Superconducting properties of mesoscopic cylinders with enhanced surface
superconductivity}
\author{B. J. Baelus, S. V. Yampolskii\cite{Sergey} and F. M. Peeters\cite{peeters}}
\address{Departement Natuurkunde, Universiteit Antwerpen (UIA), Universiteitsplein 1,%
\\
B-2610 Antwerpen, Belgium}
\author{E. Montevecchi and J. O. Indekeu}
\address{Laboratorium voor Vaste-Stoffysica en Magnetisme, Katholieke Universiteit\\
Leuven, B-3001 Leuven, Belgium}
\date{\today}
\maketitle

\begin{abstract}
The superconducting state of an infinitely long superconducting cylinder
surrounded by a medium which enhances its superconductivity near the
boundary is studied within the nonlinear Ginzburg-Landau theory. This
enhancement can be due to the proximity of another superconductor or due to
surface treatment. Quantities like the free energy, the magnetization and
the Cooper-pair density are calculated. Phase diagrams are obtained to
investigate how the critical field and the critical temperature depend on
this surface enhancement for different values of the Ginzburg-Landau
parameter $\kappa $. Increasing the superconductivity near the surface leads
to higher critical fields and critical temperatures. For small cylinder
diameters only giant vortex states nucleate, while for larger cylinders
multivortices can nucleate. The stability of these multivortex states also
depends on the surface enhancement. For type-I superconductors we found the
remarkable result that for a range of values of the surface extrapolation
length the superconductor can transit from the Meissner state into
superconducting states with vorticity $L>1$. Such a behaviour is not found
for the case of large $\kappa $, i.e. type-II superconductivity.
\end{abstract}

\pacs{74.60.De, 74.20.De, 74.25.Dw}

\section{Introduction}

The study of the superconducting properties of mesoscopic samples received
recently a lot of attention due to the progress in nanofabrication
technologies. A sample is mesoscopic if its size is comparable to the
penetration depth $\lambda $ and/or the coherence length $\xi $.

The properties of superconductors and therefore their applications are
determined by the critical parameters, i.e. the critical fields $\left( H_{c}%
\text{, }H_{c1}\text{, }H_{c2}\text{, }H_{c3}\right) $, the critical current 
$j_{c}$ and the critical temperature $T_{c}$. For application purposes one
wants to improve these parameters. Instead of searching for new bulk
materials with new physical properties, one can also try to modify the
properties of an existing material by nanostructuring the superconductor,
i.e. `quantum design'. Therefore, theoretical and experimental studies
directed to improve the critical parameters of mesoscopic superconductors
are important.

From the point of view of improving these parameters, mesoscopic thin disks
immersed in an insulating medium and placed in a perpendicular magnetic
field have attracted most attention. In disks with a small radius only giant
vortex states exist which are axially symmetric superconducting\ states \cite
{Buisson,PRL79,Benoist,PRB57}. The different giant vortex states are
characterized by their angular momentum, i.e. their vorticity $L$. For disks
with a sufficiently large radius, multivortex states can exist which are the
equivalent of the Abrikosov vortex lattice in bulk type-II superconductors $%
\left( \kappa =\lambda /\xi >1/\sqrt{2}\right) $ and the vorticity
corresponds to the number of separate vortices \cite
{Geim,PRL81,Pal1,Akkermans,SM25,PRL83,Zalalutdinov,Geim4,Pal3,PRB61,PC332p,PC332,Geim3,Saddle}%
. For such disks, it is found that the critical magnetic field depends
strongly on the size and on the Ginzburg-Landau parameter and less on the
thickness of the disk.

Fink and Presson \cite{FPR151,FPR168} studied the Meissner state and the
giant vortex state of a superconducting infinite cylinder immersed in an
insulator or vacuum and placed in a magnetic field parallel to the cylinder
axis using the Ginzburg-Landau theory. They calculated how the critical
field depends on the cylinder radius and on the Ginzburg-Landau parameter $%
\kappa $. Zharkov {\it et al} \cite{Zharkov}\ studied similar
superconducting cylinders using the Ginzburg-Landau theory and calculated
the critical fields as a function of the cylinder radius, the temperature
and the Ginzburg-Landau parameter.

Varying the sample size and the Ginzburg-Landau parameter, one can decrease
or increase the critical field, and thus the critical current, but it does
not influence the critical temperature. How can one increase the critical
temperature? Fink and Joiner \cite{FPRL23} considered a semi-infinite
superconducting half-space where the surface was treated by cold working in
such a way that the superconductivity near the surface was enhanced, i.e.
the slope of the superconducting order parameter increased near the sample
surface. They found that such a surface treatment leads to higher critical
temperatures, larger critical fields, and larger critical currents. Another
way to enhance superconductivity near the surface is by bringing the
superconductor in contact with a well chosen superconducting or
semiconducting layer \cite{Indekeu1}. The added superconducting layer must
have a higher transition temperature than the superconducting sample and in
this case the surface enhancement of superconductivity is caused by the
proximity effect. It is necessary that the semiconducting layer has a
bandgap which overlaps the superconducting gap of the superconductor.

Montevecchi and Indekeu \cite{Emma} performed a theoretical study of the
effect of confinement on the superconducting/normal transition for systems
with surface enhancement. Using the linear Ginzburg-Landau theory they
studied the critical temperature at zero field for a thin film, an infinite
cylinder and a sphere. For all geometries they found that at $H=0$ the
critical temperature increases with the enhancement of surface
superconductivity. For thin films in a parallel field, they also calculated
a $H-T$ phase diagram. They found that surface enhancement leads to an
increase of the critical field $H_{c3}$ and of the critical temperature.
Notice that their results do not depend on $\kappa ,$ since they use the
linear theory.

Yampolskii and Peeters \cite{PRB62} investigated theoretically the vortex
structure of thin mesoscopic disks in a perpendicular magnetic field
surrounded by a medium which enhances surface superconductivity. If the size
of these thin disks is sufficiently large, they found giant vortex states as
well as multivortex states appearing as metastable states and as ground
states. The enhancement of the superconductivity near the surface leads to a
stabilization of the multivortex states as ground states. They also
calculated a $H-T$ phase diagram which showed that the critical temperature
and the critical field were significantly increased by enhancing the surface
superconductivity.

In the present paper we investigate the effect of the enhancement of surface
superconductivity on the critical field and the critical temperature for
superconducting cylinders with radii equal to a few coherence lengths $\xi $%
. We also study the influence of the Ginzburg-Landau parameter $\kappa $.
Our theoretical analysis is based on a full self-consistent numerical
solution of the coupled non-linear Ginzburg-Landau equations. No a priori
arrangement of the vortex configuration or of the type of vortex
configuration is assumed.

The paper is organized as follows. In Sec. II we present our theoretical
model. We explain how the enhancement of superconductivity near the surface
is taken into account. In Sec. III we consider {\it small cylinders}, i.e.
infinitely long superconducting cylinders with small radius. In such small
cylinders only axially symmetric states or giant vortex states nucleate. We
discuss how the critical field and the critical temperature are influenced
by the surface enhancement of superconductivity. In Sec. IV {\it larger
cylinders} are studied where multivortices appear. We investigate the
dependence of the nucleation of these states on the surface enhancement. Our
results are summarized in Sec. V.

\section{Theoretical formalism}

We consider infinite superconducting cylinders with radius $R$ surrounded by
a medium which enhances superconductivity at the edge of the cylinder. Along
the axis of the cylinder a uniform magnetic field $\overrightarrow{H}_{0}$
is applied. To deal with this problem we use the Ginzburg-Landau theory and
we solve numerically and self-consistently the system of two coupled
equations 
\begin{equation}
\frac{1}{2m}\left( -i\text{{\it 
h\hskip-.2em\llap{\protect\rule[1.1ex]{.325em}{.1ex}}\hskip.2em%
}}\overrightarrow{\nabla }-\frac{2e\overrightarrow{A}}{c}\right) ^{2}\Psi
=-\alpha \Psi -\beta \Psi \left| \Psi \right| ^{2}\text{ ,}  \eqnum{1a}
\end{equation}
\begin{equation}
\overrightarrow{\nabla }\times \left( \overrightarrow{\nabla }\times 
\overrightarrow{A}\right) =\frac{4\pi }{c}\overrightarrow{j}\text{ ,} 
\eqnum{1b}
\end{equation}
where the density of superconducting current $\overrightarrow{j}$ is given
by 
\begin{equation}
\overrightarrow{j}=\frac{e\text{{\it 
h\hskip-.2em\llap{\protect\rule[1.1ex]{.325em}{.1ex}}\hskip.2em%
}}}{im}\left( \Psi ^{\ast }\overrightarrow{\nabla }\Psi -\Psi 
\overrightarrow{\nabla }\Psi ^{\ast }\right) -\frac{4e^{2}}{mc}\left| \Psi
\right| ^{2}\overrightarrow{A}\text{ .}  \eqnum{1c}
\end{equation}

The boundary condition for the vector potential is such that far away from
the cylinder the field equals the external field $H_{0}$, i.e. 
\begin{equation}
\left. \overrightarrow{A}\right| _{\overrightarrow{r}\rightarrow \infty }=%
\frac{1}{2}H_{0}\rho \overrightarrow{e}_{\phi }\text{ ,}  \eqnum{2a}
\end{equation}
where $\rho $ indicates the radial position from the center of the cylinder
and $\overrightarrow{e}_{\phi }$ is the azimuthal direction. The general
boundary condition for the order parameter can be written as \cite{Degennes} 
\begin{equation}
\left. \overrightarrow{n}\cdot \left( -i\text{{\it 
h\hskip-.2em\llap{\protect\rule[1.1ex]{.325em}{.1ex}}\hskip.2em%
}}\overrightarrow{\nabla }-\frac{2e\overrightarrow{A}}{c}\right) \right|
_{r=R}\Psi =\frac{i}{b}\left. \Psi \right| _{r=R}\text{ ,}  \eqnum{2b}
\end{equation}
where $\overrightarrow{n}$ is the unit vector normal to the cylinder and $b$
is the surface extrapolation length. The value of $b$ is determined by the
medium in which the cylinder is immersed. If the surrounding medium is
vacuum or an insulator one has $b\rightarrow \infty $, for metals $b>0$ and
for ferromagnets $b\rightarrow 0$. When, for example, the surface of the
cylinder is in contact with a superconducting layer of a higher $T_{c}$ one
has $b<0$. In the present paper we will consider the latter case of $b<0$.

Using dimensionless variables and the London gauge div$\overrightarrow{A}=0$
the GL equations can be written as 
\begin{equation}
\left( -i\overrightarrow{\nabla }-\overrightarrow{A}\right) ^{2}\Psi =\Psi
\left( 1-\left| \Psi \right| ^{2}\right) \text{ ,}  \eqnum{3a}
\label{dimeq1}
\end{equation}
\begin{equation}
-\kappa ^{2}\Delta \overrightarrow{A}=\frac{1}{2i}\left( \Psi ^{\ast }%
\overrightarrow{\nabla }\Psi -\Psi \overrightarrow{\nabla }\Psi ^{\ast
}\right) -\left| \Psi \right| ^{2}\overrightarrow{A}\text{ ,}  \eqnum{3b}
\label{dimeq2}
\end{equation}
and the boundary condition for the order parameter as 
\begin{equation}
\left. \overrightarrow{n}\cdot \left( -i\overrightarrow{\nabla }-%
\overrightarrow{A}\right) \Psi \right| _{r=R}=\left. \frac{i}{b}\Psi \right|
_{r=R}\text{ .}  \eqnum{4}
\end{equation}
Here the distance is measured in units of the coherence length $\xi $, the
order parameter in $\Psi _{0}=\sqrt{-\alpha /\beta }$, the vector potential
in $c${\it 
h\hskip-.2em\llap{\protect\rule[1.1ex]{.325em}{.1ex}}\hskip.2em%
}$/2e\xi $, and the magnetic field in $H_{c2}=c${\it 
h\hskip-.2em\llap{\protect\rule[1.1ex]{.325em}{.1ex}}\hskip.2em%
}$/2e\xi ^{2}=\kappa \sqrt{2}H_{c}$, where $H_{c}=\sqrt{4\pi \alpha
^{2}/\beta }$ is the critical field.

To solve the system of Eqs.~(\ref{dimeq1}-\ref{dimeq2}), we apply a
finite-difference representation of the order parameter and the vector
potential on a uniform Cartesian space grid (x,y), with typically 128 grid
points over a distance $2R_{o}$ (i.e. the diameter of the cylinder), and we
used the link variable approach \cite{Kato}, and an iteration procedure
based on the Gauss-Seidel technique to find $\Psi $. The vector potential is
obtained with the fast Fourier transform technique where we set $%
\overrightarrow{A}_{\left| x\right| =R_{S},\left| y\right|
=R_{S}}=H_{0}\left( x,-y\right) /2$ at the boundary of a larger space grid $%
\left( \text{typically }R_{S}=4R_{o}\right) $.

Two different vortex configurations are possible in axial symmetric samples;
the giant vortex state and the multivortex state. The giant vortex state has
cylindrical symmetry, which leads to the order parameter $\Psi =\psi \left(
\rho \right) \exp (iL\phi ),$ where $\rho $ and $\phi $ are the cylindrical
coordinates and $L$ is the angular momentum or vorticity. Along a closed
path, which lies near the edge of the cylinder, the order parameter phase
difference is given by $L$ times $2\pi $. A multivortex state is described
by a mixture of different angular harmonics which is possible due to the
nonlinearity of the GL equations. In this case the vorticity is nothing else
than the number of vortices in the superconductor.

To find the different vortex configurations, which include the metastable
states, we search for the steady-state solutions of Eqs.~(\ref{dimeq1}-\ref
{dimeq2}) starting from different randomly generated initial conditions.
Then we increase/decrease slowly the magnetic field and recalculate each
time the exact vortex structure. We do this for each vortex configuration in
a magnetic field range where the vorticity stays the same.

In order to determine which vortex state corresponds to the ground state we
calculated the difference in free energy density between the superconducting
state and the normal state 
\begin{eqnarray}
\frac{F}{F_{0}} &=&\frac{2}{V}\int_{V}dV\left[ -\left| \Psi \right| ^{2}+%
\frac{1}{2}\left| \Psi \right| ^{4}+\left| \left( -i\overrightarrow{\nabla }-%
\overrightarrow{A}\right) \Psi \right| ^{2}\right.  \nonumber \\
&&+\left. \kappa ^{2}(\overrightarrow{H}-\overrightarrow{H}_{0})^{2}\right] +%
\frac{2}{bV}\oint_{S}dS\left| \Psi \right| ^{2}\text{ ,}  \eqnum{5}
\label{energie}
\end{eqnarray}
where $\overrightarrow{H}=rot$ $\overrightarrow{A}$, $V$ is the volume of a
cylinder with unit height, $S\ $the surface of the infinite cylinder and $%
F_{0}=H_{c}^{2}V/8\pi $. The contribution of the surface is taken into
account by the last term of Eq.~(\ref{energie}). By comparing the
dimensionless free energy of the different giant and multivortex
configurations, we obtain the ground states and the metastable states. The
dimensionless magnetization is defined as 
\begin{equation}
M=\frac{\left\langle H\right\rangle -H_{0}}{4\pi }\text{ ,}  \eqnum{6}
\end{equation}
where $\left\langle H\right\rangle $ is the magnetic field averaged over the
sample. The magnetization is a direct measure of the magnetic field expelled
from the sample.

The temperature is included in $\xi $, $\lambda $, $H_{c2}$, through their
temperature dependencies 
\begin{eqnarray}
\xi (T) &=&\frac{\xi (0)}{\sqrt{\left| 1-T/T_{c0}\right| }}\text{ ,} 
\eqnum{7a} \\
\lambda (T) &=&\frac{\lambda (0)}{\sqrt{\left| 1-T/T_{c0}\right| }}\text{ ,}
\eqnum{7b} \\
H_{c2}(T) &=&H_{c2}(0)\left| 1-\frac{T}{T_{c0}}\right| \text{ ,}  \eqnum{7c}
\end{eqnarray}
where $T_{c0}$ is the critical temperature at zero magnetic field for the
normal boundary condition, i.e. $-\xi (0)/b=0$. We will only insert the
temperature explicitly if we consider the $H-T$ phase diagrams, while the
other calculations are for (arbitrary) fixed temperature. This means that we
vary $b$ at constant $\xi $ and constant $R$. Notice further that the
Ginzburg-Landau parameter $\kappa =\lambda /\xi $ is independent of the
temperature.

\section{Small cylinders}

First we discuss infinitely long cylinders with a small radius, where we may
limit ourselves to the giant vortex states. In this case the confinement
effects are dominant and they impose the cylindrical symmetry of the
boundary of the cylinder on the vortex state. Consequently, the dimensions
of the Ginzburg-Landau equations are reduced, which improves the accuracy
and the computation time.

\subsection{Type-I cylinders}

We consider infinite long cylinders with radius $R=2.0\xi $ for a
Ginzburg-Landau parameter $\kappa =0.28$.

Figs.~\ref{emagR2k028} (a,b) show the {\it ground state} free energy, i.e.
the thermodynamic equilibrium free energy, and the magnetization as a
function of the applied magnetic field for such a superconducting cylinder
for the usual boundary condition, i.e. $-\xi /b=0.0$ (solid curves), and for 
$-\xi /b=0.2$ (dashed curves) and $0.4$ (dash-dotted curves), which
correspond to surface enhancement. With increasing $-\xi /b$ the free energy
at zero magnetic field becomes more negative, which means an enhancement of
superconductivity. Also the ground state superconducting/normal transition
moves to higher fields. Notice that the magnetic field expulsion (see Fig.~%
\ref{emagR2k028}(b)) is enhanced with increasing $-\xi /b$.

Figs.~\ref{coopR2k028}(a-c) show the radial dependence of the Cooper-pair
density, the magnetic field and the current density, respectively, for the
previously considered cylinder at the applied magnetic field $%
H_{0}/H_{c2}=2.02$ and for vorticity $L=0$. The solid curves give the result
for $-\xi /b=0$, the dashed curves for $-\xi /b=0.2$ and the dash-dotted
curves for $-\xi /b=0.4$. From Fig.~\ref{coopR2k028}(a) it is very clear
that the Cooper-pair density near the surface, and hence the surface
superconductivity, increases with increasing $-\xi /b$. For a small cylinder
radius, also the Cooper-pair density in the center of the cylinder is
influenced by the boundary condition. From Fig.~\ref{coopR2k028}(b) one can
see that enhancing the surface superconductivity results in a more
pronounced Meissner effect, i.e. the magnetic field expulsion becomes more
complete with increasing $-\xi /b$. Notice that the magnetic field for this
situation is always zero in the center of the cylinder. To expel the
magnetic field more, the superconductor has to induce more superconducting
current near the surface. Fig.~\ref{coopR2k028}(c) shows that the current
density is zero in the center of the superconductor and becomes more
negative near the surface. The superconducting Meissner currents are also
more concentrated near the surface with increasing $-\xi /b.$ For the
Meissner state the current has the same sign in the whole sample and is thus
flowing in the whole sample in the same direction.

In Fig.~\ref{hbR2k028} the dependence of the transition fields for the
different giant vortex states on the surface enhancement $b$ is given for a
superconducting cylinder with radius $R=2.0\xi $ and $\kappa =0.28$. The
solid curves give the ground state transitions between the different $L$%
-states and the thick solid curve indicates the superconducting/normal
transition. For the normal boundary condition, $-\xi /b=0$, it is evident
that the ground state consists of the Meissner state, i.e. $L=0$, since $%
\kappa $ is sufficiently small. The superconducting/normal transition occurs
at $H_{0}/H_{c2}\approx 2.9$. What happens with increasing $-\xi /b$? (i)
For $-\xi /b<0.65$ the ground state is still given by the Meissner state,
and with increasing $-\xi /b$ the superconducting/normal-transition moves to
higher fields. (ii) For $-\xi /b>0.65$ different $L$ states become the
ground state. This is in agreement with the results of Montevecchi \cite
{Emmaphd} who also found transitions between different $L$ states for $1/b<0$
for a fixed cylinder radius $R/\left| b\right| =1.2$ and $\kappa =0.3$ (see
Fig.~1.14 of Ref.~\cite{Emmaphd}). Remarkably we find that the ground state
does not evolve from the Meissner state to a state with vorticity $L=1$.
Rather it evolves to a state with larger vorticity. For $0.65\lesssim -\xi
/b\lesssim 0.8$ the ground state changes from the Meissner state to a state
with $L=4$, for $0.8\lesssim -\xi /b\lesssim 1.8$ to a state with $L=3,$ and
for $-\xi /b\gtrsim 1.8$ to a state with $L=2$. (iii) Further increasing $%
-\xi /b$ the ground state will change first from $L=0$ to $L=1$. The latter
transition corresponds to type-II behavior which only occurs in the absence
of surface enhancement for $\kappa $ larger than some critical
Ginzburg-Landau parameter $\kappa _{2}$.

The three different regimes which we found for the case of increasing $-\xi
/b$ for fixed small $\kappa $ are very similar to the three regimes for the
case of increasing $\kappa $ for the normal boundary condition, i.e. $-\xi
/b=0$: (i) For the normal boundary condition it is known \cite{Roseinnes}
that only the Meissner state is the ground state for $\kappa $ smaller than
a critical parameter $\kappa _{1}$. (ii) For $\kappa _{1}\lesssim \kappa
\leq \kappa _{2}$ surface superconductivity can occur and states with
vorticity larger than one exist. For example, Fink and Presson \cite{FPR151}
found ground state transitions from the Meissner state with $L=0$ into
surface superconducting states with $L=4$ for a superconducting cylinder
with radius $R=3.0\xi $ and $\kappa =0.5$. (iii) For $\kappa >\kappa _{2}$
(type II superconductivity) the ground state transits from the Meissner
state to states with $L=1$, $2$, ..., $n$, successively, where the value of $%
n$ depends on the radius and the Ginzburg-Landau parameter (see for example
Ref.~\cite{FPR151}). In bulk superconductors $\kappa _{1}\approx 0.42$ and $%
\kappa _{2}=1/\sqrt{2}$, but for cylinders these parameters are radius
dependent \cite{Zharkov}.

What is the reason for this remarkable behaviour of the ground state for
increasing $-\xi /b$? (i) For $-\xi /b<0.65$ the order parameter, and thus
the Cooper-pair density, near the cylinder boundary increases with
increasing $-\xi /b$ which also leads to an enhancement of the order
parameter in the center, albeit less pronounced (see also Fig.~\ref
{coopR2k028}(a)). Consequently, the free energy at zero magnetic field
becomes more negative and the superconducting/normal transition moves to
higher fields (see also Fig.~\ref{emagR2k028} (a)). (ii) With increasing $%
-\xi /b$ the Cooper-pair density in the center for the $L=0$ state is not
increasing as strongly as the Cooper-pair density near the boundary and the
cylinder remains superconducting at higher fields. For $-\xi /b\gtrsim 0.65$
it becomes energetically less favourable to expell rather high fluxes from
the sample. The effect of the increase of vorticity on the free energy is
shown in Fig.~\ref{cb07} for $-\xi /b=0.7$. From the inset it is clear that
the free energy of the $L=4$ state becomes lower than the one of the
Meissner state at $H_{0}/H_{c2}\approx 5.12.$ This means that for this field
the penetration of 4 fluxes becomes more favourable than the expulsion of
the field. At $H_{0}/H_{c2}\approx 5.20$, $5.30$ and $5.32$ the ground state
changes respectively into the $L=5$ state, the $L=6$ state and the normal
state. Further increasing $-\xi /b$ leads to higher Cooper-pair density near
the boundary and, consequently, to higher induced supercurrents. As a
result, a smaller amount of flux will penetrate into the sample after the
first ground state transition. For example, with increasing field at $-\xi
/b>0.8$ the ground state changes from the Meissner state into the $L=3$
state and at higher fields into states with $L=4$, $5$, $6$ and so on. (iii)
For high values of $-\xi /b,$ the explanation is analogous as above but now
the ground state changes from $L=0$ to $L=1$, $2$, $3$ and so forth.

In Fig.~\ref{htR2k028} we plot the $H-T$ phase diagram for a superconducting
cylinder with $R=2.0\xi $ and $\kappa =0.28$ for $-\xi (0)/b=0.0$, $0.1$ and 
$0.2$. For these values of $-\xi (0)/b$ the superconducting ground state is
always given by the Meissner state (see also Fig.~\ref{hbR2k028}). The
superconducting/normal transition is shown by solid curves for $-\xi
(0)/b=0.0$, by dashed curves for $-\xi (0)/b=0.1$, and by dash-dotted curves
for $-\xi (0)/b=0.2$. For $\kappa =0$ the superconducting/normal transition
is a straight line in the $H-T$ phase diagram. For $\kappa =0.28$, the
superconducting/normal transition is still given by a straight line for $%
T\ll T_{c}$, regardless of the value of $-\xi (0)/b$, but it has now a
curvature near $T_{c}$. With increasing $-\xi (0)/b$ the
superconducting/normal transition, which is of first order, moves to higher
temperatures for fixed field or to higher fields for fixed temperature.

\subsection{Type-II cylinders}

We consider infinite cylinders with radius $R=2.0\xi $ for a Ginzburg-Landau
parameter $\kappa =1.0$.

Figs.~\ref{emagR2k100}(a,b) show the {\it ground state} free energy and
magnetization as a function of the applied magnetic field for such a
superconducting cylinder for $-\xi /b=0.0$ (solid curves), $0.2$ (dashed
curves) and $0.4$ (dash-dotted curves). Now, the superconducting ground
state is not always the Meissner state. For example, for the usual boundary
condition, i.e. $-\xi /b=0.0$, the vorticity of the ground state is $L=0$
for $H_{0}/H_{c2}\leq 1.195$. At the first transition field $%
H_{0}/H_{c2}=1.195$ the vorticity of the ground state changes from $L=0$ to $%
L=1$ and then it remains $L=1$ until $H_{0}/H_{c2}=1.7575$ where it changes
to $L=2$. At $H_{0}/H_{c2}=2.12$ the free energy of the ground state becomes
zero and the superconducting/normal transition takes place. The transitions
between the different $L$-states are indicated by corners in the free energy
and jumps in the magnetization. As for type-I cylinders, the free energy at $%
H_{0}=0$ becomes more negative with increasing $-\xi /b$, i.e.
superconductivity is enhanced, and the transition to the normal state moves
to higher fields. Furthermore, the vorticity of the ground state can become
larger and the peaks in the magnetization are higher, indicating a more
efficient expulsion of the magnetic field from the superconductor. Notice
that the magnetic fields for the transitions between different $L$-states
are almost independent of the value of $-\xi /b$, but the surface critical
field $H_{c3}$ is a sensitive function of $-\xi /b$.

Next, we investigate the radial dependence of the Cooper-pair density, the
magnetic field and the current density for such a superconducting cylinder.
For vorticity $L=0$ the results are analogous to the results for type-I
cylinders which were described in Fig.~\ref{coopR2k028}. In Figs.~\ref
{coopR2k100}(a-c) we show the radial dependence of, respectively, the
Cooper-pair density, the magnetic field and the current density for the
ground state at $H_{0}/H_{c2}=2.02$, i.e. the giant vortex state with
vorticity $L=2$. The solid curves give the result for $-\xi /b=0$, the
dashed curves for $-\xi /b=0.2$ and the dash-dotted curves for $-\xi /b=0.4$%
. In Fig.~\ref{coopR2k100}(a) it is shown that the Cooper-pair density is
zero for all $-\xi /b$ in the center of the cylinder, i.e. at the position
of the giant vortex with vorticity $L=2$. Near the surface the Cooper-pair
density is enhanced, which is understandable because the slope of $\left|
\Psi \right| $ at $\rho =R$ is given by $-\xi /b$. For the giant vortex
state, the magnetic field is not zero in the center, as was the case for $%
L=0 $, but it can be even higher than the external field $H_{0}=2.02H_{c2}$.
With increasing $-\xi /b$, the superconductivity near the boundary is
enhanced and, thus, more magnetic field can be expelled from the cylinder
and the giant vortex will be more compressed in the center (see Fig.~\ref
{coopR2k100}(b)). As a consequence, the minimum of magnetic field inside the
cylinder decreases and the magnetic field at the position of the vortex
increases. To expel the magnetic field more and to compress the giant vortex
better in the center, the superconductor has to induce a larger
superconducting current. Therefore the superconducting current is more
positive close to the giant vortex and more negative near the boundary (see
Fig.~\ref{coopR2k100}(c)).

The fact that the transition fields between different $L$-states are almost
independent of the value of $-\xi /b$, can also be seen from the $\left(
-\xi /b\right) -H$ phase diagram for a superconductor cylinder with radius $%
R=2.0\xi $ and $\kappa =1.0$, which is shown in Fig.~\ref{hbR2k100}. The
thin solid curves give the ground state transitions between the different $L$%
-states and the thick solid curve indicates the superconducting/normal
transition. For $\kappa =1.0$ we do not find ground state transitions
between a state with vorticity $L=0$ and vorticity $L>1$, as was the case
for $\kappa =0.28$. Moreover, the magnetic field range over which the ground
state has a particular vorticity $L$ is almost the same for all $L>0$,
namely $\Delta H_{L}=H_{L\rightarrow L+1}-H_{L-1\rightarrow L}\approx
0.57H_{c2}$. Fig.~\ref{coopR2k100} indicates that the width of the
superconducting sheath in $\rho /\xi $ is independent of $b$. The value of $%
b $ influences only the amplitude of the order parameter near the surface
(Fig.~\ref{coopR2k100}(a)), the diamagnetization (Fig.~\ref{coopR2k100}(b))
and the strength of the currents (Fig.~\ref{coopR2k100}(c)). Therefore, the
space available for the confined flux in the core region of the cylinder
does (almost) not depend on the enhancement. As a consequence, the number of
fluxoids, or vorticity $L$, depends only on the strength of the external
field $H_{0}$ and almost not on $b$ for fixed temperature.

In Fig.~\ref{htR2k100} we plot the $H-T$ phase diagram for a superconducting
cylinder with $R=2.0\xi $ and $\kappa =1.0$ for $-\xi (0)/b=0.0$, $0.1$ and $%
0.2$. The ground state transitions are shown by solid curves for $-\xi
(0)/b=0.0$, by dashed curves for $-\xi (0)/b=0.1$, and by dash-dotted curves
for $-\xi (0)/b=0.2$. The thick curves indicate the superconducting/normal
transition and the thinner curves (they are almost straight lines) show the
ground state transitions between the giant vortex states $L\leftrightarrow
L+1$. From Fig.~\ref{htR2k100} it is clear that the transitions between the
different $L$ states are (almost) independent of the value of $-\xi /b$, but
the superconducting/normal transition is a sensitive function of $-\xi /b$
(see also Fig.~\ref{hbR2k100}). With increasing $-\xi (0)/b$ the
superconducting/normal transition moves to higher temperatures for fixed
field or to higher fields for fixed temperature as was also the case for $%
\kappa =0.28$. The zero field critical temperature is the same for $\kappa
=1.0$ as for $\kappa =0.28$ and depends only on $-\xi (0)/b$. The critical
fields at zero temperature are different in both cases. For $-\xi (0)/b=0.2,$
for example, $H_{c3}(0)=3.5H_{c2}(0)$ for $\kappa =0.28$ and $%
H_{c3}(0)=2.7H_{c2}(0)$ for $\kappa =1.0$. In fact, the `zero' temperature
corresponds to some reference temperature rather than the real temperature
because the Ginzburg-Landau theory is not valid at very low temperatures.
Another significant difference is that for $\kappa =1.0$ the
superconducting/normal transition is no longer a straight line for
temperatures $T\ll T_{c}$, but consists of corners which indicate the
transition between the different $L$ states. These corners are {\it %
bicritical points }where three phases coexist for a given value of $-\xi
(0)/b$. For example, the normal state coexists with the $L=0$ state and the $%
L=1$ state at $H_{0}/H_{c2}(0)=0.86$ and $T/T_{c0}=0.84$ for $-\xi (0)/b=0.2$%
. Note that all transitions between different $L$ states are of first order
and correspond to free energy crossings, while the superconducting/normal
transition is a second order transition. These bicritical points occur at
magnetic fields which are only weakly sensitive to the surface enhancement,
while the transition temperatures are quite sensitive to $b$. Notice further
that the superconducting/normal phase transition boundary is exactly the
same for a cylinder with $\kappa =1.0$ as for a very thin disk (thickness $%
d\ll \xi $) with the same radius, if the same boundary condition (Eq.~(2b))
is taken on the sides, but the standard boundary condition ($-\xi /b=0$) on
the top and the bottom of the disk. This leads to a $z$-independent order
parameter and, as a consequence, to similar physics as for an infinitely
long cylinder. This can be seen by comparing Fig.~\ref{htR2k100} with
Fig.~14 of Ref.~\cite{PRB62}. In the case of the disk, the field is
perpendicular to the disk.

\section{Large cylinders: multivortices}

Up to now, we considered only cylinders with small radii where the
confinement effects were dominant and only giant vortex states were stable.
Now, we consider superconducting cylinders with a large radius in which
multivortex states can nucleate for certain magnetic fields. This means that
we can no longer limit our calculations to axially symmetric solutions.
Nevertheless, from the study of disks \cite{PRL81} and rings \cite{PRB61} we
know that the transition fields between states with different vorticity do\
(almost) not depend on the fact whether one considers axial symmetry or not.
Therefore, we can still calculate the $\left( -\xi /b\right) -H$ phase
diagrams using the cylindrical symmetry approach, but if we want to know the
real vortex configurations, we have to use the general theory in order to
account for multivortices.

We calculate the $\left( -\xi /b\right) -H$ phase diagrams for a
superconducting cylinder with radius $R=4.0\xi $ for $\kappa =0.28$ and $%
\kappa =1.0$, which are shown in Figs.~\ref{hbR4}(a,b), respectively. The
thin solid curves give the ground state transitions between the different $L$%
-states and the thick solid curve indicates the superconducting/normal
transition. The conclusions are similar to those for small cylinders, but
with many more transitions between different $L$-states. However, for $%
\kappa =0.28$ we find now transitions from the Meissner state immediately to
a state with vorticity $L=13$ as a function of the magnetic field in the
range $0.43\leq -\xi /b\leq 0.47$. With further increasing $-\xi /b$ we find
first transitions from $L=0$ to $L=12$, $11$, $10$ and $9$, respectively.
Notice that {\it triple points} occur where the Meissner state coexists with
two different giant vortex states. For example, at $H_{0}/H_{c2}\approx 3.4$
and $-\xi /b\approx 0.47$ the Meissner state coexists with the $L=12$ state
and the $L=13$ state. This means that these three states have the same free
energy in that case.

For $\kappa =1.0$ the magnetic field range over which the ground state has a
particular vorticity $L$ is almost the same for all $L>0$, namely $\Delta
H_{L}\approx 0.125H_{c2}$, which means that the ground state changes more
quickly from vorticity $L$ than for cylinders with radius $R=2.0\xi $, where 
$\Delta H_{L}\approx 0.57H_{c2}$, which is due to the larger cross section
of the cylinder.

Next we will compare the free energy and the magnetization for the usual
boundary condition $-\xi /b=0$ with the case of surface enhancement, $-\xi
/b=0.2$. Figs.~\ref{emagR4cb00}(a,b) show the free energy and the
magnetization for a cylinder with radius $R=4.0\xi $ and $\kappa =1.0$ for
the usual boundary condition. The different giant vortex states are
indicated by the solid curves, the multivortex states by the dashed curves
and the multivortex to giant vortex transition fields $H_{MG}$ by the open
circles. Notice that we find superconducting states up to vorticity $L=11$
and that the superconducting/normal transition field is $H_{c3}/H_{c2}%
\approx 1.85$ (see also Fig.~\ref{hbR4}(b)). Multivortices can nucleate for
vorticities $L=2$, $3$, and $4$. The multivortex state with $L=2$ can
nucleate for $0.295\leq H_{0}/H_{c2}\leq 0.92$, with $L=3$ for $0.42\leq
H_{0}/H_{c2}\leq 1.045$, and with $L=4$ for $0.5325\leq H_{0}/H_{c2}\leq
0.92 $. In Fig.~\ref{emagR4cb00}(b) the equilibrium ground state transitions
are indicated by the vertical curves.

Figs.~\ref{emagR4cb02}(a,b) show the free energy and the magnetization for a
cylinder with radius $R=4.0\xi $ and $\kappa =1.0$ for $-\xi /b=0.2.$ The
different giant vortex states are indicated by the solid curves, the
multivortex states by the dashed curves and the multivortex to giant vortex
transition fields $H_{MG}$ by the open circles. The latter transition is of
second order.\ For this boundary condition the superconducting/normal
transition field is $H_{c3}/H_{c2}\approx 2.35$ which is appreciably higher
than for $-\xi /b=0$. Also vortex states with higher vorticity are stable,
up to 15 instead of up to 11 for $-\xi /b=0$. Multivortex states can
nucleate with $L=2$ for $0.1325\leq H_{0}/H_{c2}\leq 0.945$, with $L=3$ for $%
0.2575\leq H_{0}/H_{c2}\leq 1.02$, and with $L=4$ for $0.3825\leq
H_{0}/H_{c2}\leq 0.895$.

How does the value of $-\xi /b$ influence the nucleation of the multivortex
states? Fig.~\ref{multistab} shows the free energy of the $L=3$ state for a
superconducting cylinder with radius $R=4.0\xi $ and $\kappa =1.0$ for $-\xi
/b=0.0$, $0.2$, $0.4$, $0.6$, $0.8$ and $1.0$ as a function of the applied
magnetic field. Giant vortex states are given by solid curves and
multivortex states by dashed curves. The transitions from a multivortex
state to a giant vortex state are indicated by the open circles. With
increasing $-\xi /b$ the $L=3$ state can nucleate up to larger fields, i.e. $%
H_{nuc}/H_{c2}=1.21$, $1.32$, $1.47$, $1.66$, $1.88$ and $2.16$ for $-\xi
/b=0.0$, $0.2$, $0.4$, $0.6$, $0.8$ and $1.0$, respectively. On the other
hand the multivortex to giant vortex transition field decreases, i.e. $%
H_{MG}/H_{c2}=1.045$, $1.02$, $1.0075$, $0.995$, $0.97$ and $0.82$ for $-\xi
/b=0.0$, $0.2$, $0.4$, $0.6$, $0.8$ and $1.0$, respectively. This means that
with increasing $-\xi /b$ the magnetic field region over which the giant
vortex state exists increases and the one of the multivortex state
decreases. The reason is that increasing $-\xi /b$ corresponds to an
enhancement of superconductivity near the boundary. Therefore, it is more
difficult for the magnetic field to penetrate near the cylinder edge than in
the center. Since multivortices are situated more closely to the cylinder
edge, it is obvious that multivortices become unstable at lower fields for
increasing $-\xi /b$. Or alternatively, the confinement effects become more
important for increasing surface superconductivity.

To illustrate this better, we show the Cooper-pair density of the $L=3$
state for such a cylinder in Figs.~\ref{multidens}(a-f) at $%
H_{0}/H_{c2}=0.445$ for $-\xi /b=0.0$, $0.2$, $0.4$, $0.6$, $0.8$ and $1.0$,
respectively. High Cooper-pair density is given by dark regions, low
Cooper-pair density by light regions. The three vortices correspond to the
three white spots. For the usual boundary condition $-\xi /b=0.0$ the
vortices are clearly separated. With increasing $-\xi /b$ the vortices are 
{\it pushed} to the center and at $-\xi /b=0.4$ they already start to
overlap. For $-\xi /b=0.8$ and $10$ the vortices are very close to each
other. Therefore, we plotted the logarithm of the Cooper-pair density to
show the positions of the vortices and to prove that it it still a
multivortex state. Thus, by the enhancement of the Cooper-pair density near
the boundary the vortices move to the center and for increasing $-\xi /b$
they will recombine in the center creating a giant vortex state.

\section{Conclusions}

In the present paper we investigated the effect of the enhancement of
surface superconductivity on the critical field and the critical temperature
for superconducting cylinders with radii comparable to the coherence length $%
\xi $. We also studied the influence of a the Ginzburg-Landau parameter $%
\kappa $. A distinction was made between cylinders with small radii where
the confinement effects dominate and only giant vortex states exist, and
cylinders with a larger radius where multivortices can nucleate for certain
magnetic fields and vorticities.

Generally, increasing $-\xi /b$ leads to a more negative free energy at $H=0$
and to a higher superconducting/normal transition field. We also studied the
magnetic field distribution, the Cooper-pair density and the current density
for different values of the surface enhancement. For higher values of $-\xi
/b$ more magnetic field can be expelled from the cylinder and the giant
vortex in the center is compressed more. Therefore, higher currents are
induced near the boundary and near the giant vortex. The Cooper-pair density
close to the boundary increases, and for small cylinders this also
influences the Cooper-pair density in the center.

From the $H-T$ phase diagrams we found that the critical temperature depends
on $-\xi /b$, while it is independent of $\kappa $ (cf. Ref.~\cite{Emma}).
With increasing $-\xi /b$ the critical temperature increases for fixed
magnetic field and the critical magnetic field increases for fixed
temperature.

We also obtained $\left( -\xi /b\right) -H$ phase diagrams. In type-I
cylinders the surface enhancement has drastic consequences. Even at low $%
\kappa $ the surface enhancement leads to transitions between different $L$%
-state and thus to type-II behavior. Moreover, as a function of the magnetic
field the superconducting ground state transits from the Meissner state to a
vortex state with $L>1$ over a range of $-\xi /b$ values. Therefore, we can
conclude that increasing $-\xi /b$ for type-I superconductors seems to have
a similar effect as increasing $\kappa $ for fixed $-\xi /b$ for certain
properties of the superconducting cylinder. In type-II cylinders we found
that the magnetic field range over which the ground state has a particular
vorticity $L$ is almost the same for all $L>1$. This magnetic field range
decreases with increasing cylinder radius.

If the cylinder radius is sufficiently large multivortex states can nucleate
and we studied the influence of the surface enhancement on the nucleation of
these states. We found that at fixed field and with increasing $-\xi /b$ the
multivortices move to the center creating a giant vortex state. Thus surface
enhancement destabilizes the multivortex state.

\section{Acknowledgments}

This work was supported by the Flemish Science Foundation (FWO-Vl) Project
No. G.0277.97N, the Flemish Concerted Action (GOA), the Inter-University
Attraction Poles (IUAP) research programs, and the European ESF-Vortex
Matter. Discussions with G. F. Zharkov are gratefully acknowledged.

\begin{figure}[tbp]
\caption{(a) The ground free energy and (b) the magnetization as a function
of the applied magnetic field for a superconducting cylinder with radius $%
R=2.0\protect\xi $ for $\protect\kappa =0.28$ and $-\protect\xi /b=0$ (solid
curves), $0.2$ (dashed curves) and $0.4$ (dash-dotted curves).}
\label{emagR2k028}
\end{figure}

\begin{figure}[tbp]
\caption{(a) The Cooper-pair density, (b) the magnetic field, and (c) the
current density as a function of the radial position for a cylinder with
radius $R=2.0\protect\xi $ and for $\protect\kappa =0.28$. The external
field is $H_{0}/H_{c2}=2.02$ and the vorticity is $L=0$.}
\label{coopR2k028}
\end{figure}

\begin{figure}[tbp]
\caption{The $\left( -\protect\xi /b\right) -H$ phase diagram for a
superconducting cylinder with radius $R=2.0\protect\xi $ and $\protect\kappa %
=0.28$. The solid curves indicate the transitions between the different $L$%
-states and the thick curve gives the superconducting/normal transition.}
\label{hbR2k028}
\end{figure}

\begin{figure}[tbp]
\caption{The free energy for a superconducting cylinder with radius $R/%
\protect\xi =2.0$ and $\protect\kappa =0.28$ for $-\protect\xi /b=0.7$ as a
function of the external magnetic field. }
\label{cb07}
\end{figure}

\begin{figure}[tbp]
\caption{The $H-T$ phase diagram for a superconducting cylinder with radius $%
R=2.0\protect\xi $ and $\protect\kappa =0.28$ for $-\protect\xi /b=0.0$
(solid curves), $0.2$ (dashed curves) and $0.4$ (dash-dotted curves).}
\label{htR2k028}
\end{figure}

\begin{figure}[tbp]
\caption{(a) The ground state free energy and (b) the magnetization as a
function of the applied magnetic field for a superconducting cylinder with
radius $R=2.0\protect\xi $ for $\protect\kappa =1.0$ and $-\protect\xi /b=0$
(solid curves), $0.2$ (dashed curves) and $0.4$ (dash-dotted curves).}
\label{emagR2k100}
\end{figure}

\begin{figure}[tbp]
\caption{(a) The Cooper-pair density, (b) the magnetic field, and (c) the
current density as a function of the radial position for a cylinder with
radius $R=2.0\protect\xi $ and for $\protect\kappa =1.0$. The external field
is $H_{0}/H_{c2}=2.02$ and the vorticity is $L=2$.}
\label{coopR2k100}
\end{figure}

\begin{figure}[tbp]
\caption{The $\left( -\protect\xi /b\right) -H$ phase diagram for a
superconducting cylinder with radius $R=2.0\protect\xi $ and $\protect\kappa %
=1.0$. The solid curves indicate the transitions between the different $L$%
-states and the thick curve gives the superconducting/normal transition.}
\label{hbR2k100}
\end{figure}

\begin{figure}[tbp]
\caption{The $H-T$ phase diagram for a superconducting cylinder with radius $%
R=2.0\protect\xi $ and $\protect\kappa =1.0$ for $-\protect\xi /b=0.0$
(solid curves), $0.2$ (dashed curves) and $0.4$ (dash-dotted curves). The
thick curves indicate the superconducting/normal transition and the thinner
curves the ground state transitions between the giant vortex states with
different vorticity $L$.}
\label{htR2k100}
\end{figure}

\begin{figure}[tbp]
\caption{The $\left( -\protect\xi /b\right) -H$ phase diagram for a
superconducting cylinder with radius $R=4.0\protect\xi $ for $\protect\kappa %
=0.28$ (a) and $\protect\kappa =1.0$ (b). The solid curves indicate the
transitions between the different $L$-states and the thick curve gives the
superconducting/normal transition.}
\label{hbR4}
\end{figure}

\begin{figure}[tbp]
\caption{(a) Free energy and (b) magnetization for vortex states in a
superconducting cylinder with radius $R=4.0\protect\xi $ and $\protect\kappa %
=1.0$ for the normal boundary condition $-\protect\xi /b=0$. Giant vortex
states are given by solid curves, multivortex states by dashed curves and
the open circles indicate the multivortex to giant vortex transition fields.
The vertical lines in (b) give the ground state transitions.}
\label{emagR4cb00}
\end{figure}

\begin{figure}[tbp]
\caption{(a) Free energy and (b) magnetization for vortex states in a
superconducting cylinder with radius $R=4.0\protect\xi $ and $\protect\kappa %
=1.0$ for the normal boundary condition $-\protect\xi /b=0.2$. Giant vortex
states are given by solid curves, multivortex states by dashed curves and
the open circles indicate the multivortex to giant vortex transition fields.
The vertical lines in (b) give the ground state transitions.}
\label{emagR4cb02}
\end{figure}

\begin{figure}[tbp]
\caption{Free energy for the $L=3$ state in a superconducting cylinder with
radius $R=4.0\protect\xi $ and $\protect\kappa =1.0$ for $-\protect\xi
/b=0.0 $, $0.2$, $0.4$, $0.6$, $0.8$ and $1.0.$ Giant vortex states are
given by solid curves, multivortex states by dashed curves and the open
circles indicate the multivortex to giant vortex transition fields.}
\label{multistab}
\end{figure}

\begin{figure}[tbp]
\caption{The Cooper-pair density for the multivortex state with $L=3$ in a
cylinder with $R/\protect\xi =4.0$ and $\protect\kappa =1.0$ at $%
H_{0}/H_{c2}=0.445$ for $-\protect\xi /b=0.0$ (a), $0.2$ (b), $0.4$ (c), $%
0.6 $ (d), and the logarithm of the Cooper-pair density for $-\protect\xi
/b=0.8$ (e) and $1.0$ (f).}
\label{multidens}
\end{figure}


\begin{references}
\bibitem[*]{Sergey}  Permanent address: Donetsk Physical \& Technical
Institute, National Academy of Sciences of Ukraine, Donetsk 83114, Ukraine.

\bibitem[%
\circ%
%
]{peeters}  Electronic mail: peeters@uia.ua.ac.be

\bibitem{Buisson}  O. Buisson, P. Gandit, R. Rammel, Y. Y. Wang, and B.
Pannetier, Phys. Lett. A {\bf 150}, 36 (1990).

\bibitem{PRL79}  P. S. Deo, V. A. Schweigert, F. M. Peeters, and A. K. Geim,
Phys. Rev. Lett. {\bf 79}, 4653 (1997).

\bibitem{Benoist}  R. Benoist and W. Zwerger, Z. Phys. B {\bf 103}, 377
(1997).

\bibitem{PRB57}  V. A. Schweigert and F. M. Peeters, Phys. Rev. B {\bf 57},
13817 (1998).

\bibitem{Geim}  A. K. Geim, I. V. Grigorieva, S. V. Dubonos, J. G. S. Lok,
J. C. Maan, A. E. Filippov, and F. M. Peeters, Nature (London) {\bf 390},
256 (1997).

\bibitem{PRL81}  V. A. Schweigert and F. M. Peeters, and P. S. Deo, Phys.
Rev. Lett. {\bf 81}, 2783 (1998).

\bibitem{Pal1}  J. J. Palacios, Physica B {\bf 256-258}, 610 (1998); {\it %
ibid.} Phys. Rev. B {\bf 58}, R5948 (1998).

\bibitem{Akkermans}  E. Akkermans and K. Mallick, J. Phys. A {\bf 32}, 7133
(1999).

\bibitem{SM25}  P. S. Deo, F. M. Peeters, and V. A. Schweigert, Superlatt.
and Microstruct. {\bf 25}, 1195 (1999).

\bibitem{PRL83}  V. A. Schweigert and F. M. Peeters, Phys. Rev. Lett. {\bf 83%
}, 2409 (1999).

\bibitem{Zalalutdinov}  M. Zalalutdinov, H. Fujioka, Y. Hashimoto, S.
Katsumoto, and Y. Iye, J. Phys. Soc. Jpn. {\bf 68}, 2872 (1999).

\bibitem{Geim4}  A. K. Geim, S. V. Dubonos, J. J. Palacios, I. V.
Grigorieva, M. Henini, and J. J. Schermer, Phys. Rev. Lett. {\bf 85}, 1528
(2000).

\bibitem{Pal3}  J. J. Palacios, Phys. Rev. Lett. {\bf 84}, 1796 (2000).

\bibitem{PRB61}  B. J. Baelus, F. M. Peeters, and V. A. Schweigert, Phys.
Rev. B {\bf 61}, 9734 (2000).

\bibitem{PC332p}  F. M. Peeters, V. A. Schweigert, B. J. Baelus, and P. S.
Deo, Physica C {\bf 332}, 255 (2000).

\bibitem{PC332}  V. A. Schweigert and F. M. Peeters, Physica C {\bf 332},
426 (2000).

\bibitem{Geim3}  A. K. Geim, S. V. Dubonos, I. V. Grigorieva, K. S.
Novoselov, F. M. Peeters, and V. A. Schweigert, Nature (London) {\bf 407},
55 (2000).

\bibitem{Saddle}  B. J. Baelus, F. M. Peeters, and V. A. Schweigert, Phys.
Rev. B {\bf 63}, 144517 (2001).

\bibitem{FPR151}  H. J. Fink and A. G. Presson, Phys. Rev. {\bf 151}, 219
(1966).

\bibitem{FPR168}  H. J. Fink and A. G. Presson, Phys. Rev. {\bf 168}, 399
(1968).

\bibitem{Zharkov}  G. F. Zharkov, V. G. Zharkov, and A. Yu. Zvetkov, Phys.
Rev. B {\bf 61}, 12293 (2000).

\bibitem{FPRL23}  H. J. Fink and W. C. H. Joiner, Phys. Rev. Lett. {\bf 23},
120 (1969).

\bibitem{Indekeu1}  J. O. Indekeu, F. Clarysse, and E. Montevecchi, to
appear in the Proceedings of the NATO ASI, Albena, Bulgaria (1998).

\bibitem{Emma}  E. Montevecchi and J. O. Indekeu, Phys. Rev. B {\bf 62},
14359 (2000), {\it ibid.} Europhys. Lett. {\bf 51}, 661 (2000).

\bibitem{PRB62}  S. V. Yampolskii and F. M. Peeters, Phys. Rev. B {\bf 62},
9663 (2000).

\bibitem{Degennes}  P. G. de Gennes, {\it Superconductivity of Metals and
Alloys }(Addison-Wesley, New York, 1994).

\bibitem{Kato}  R. Kato, Y. Enomoto, and S. Maekawa, Phys. Rev. B {\bf 47},
8016 (1993).

\bibitem{Emmaphd}  E. Montevecchi, {\it Theory of confinement effects on
superconductivity}, Ph. D. Thesis, Katholieke Universiteit Leuven, Leuven
(2000).

\bibitem{Roseinnes}  A. C. Rose-Innes and E. H. Rhoderick, {\it Introduction
to Superconductivity} (Pergamon Press, Oxford, 1976).\bigskip
\end{references}
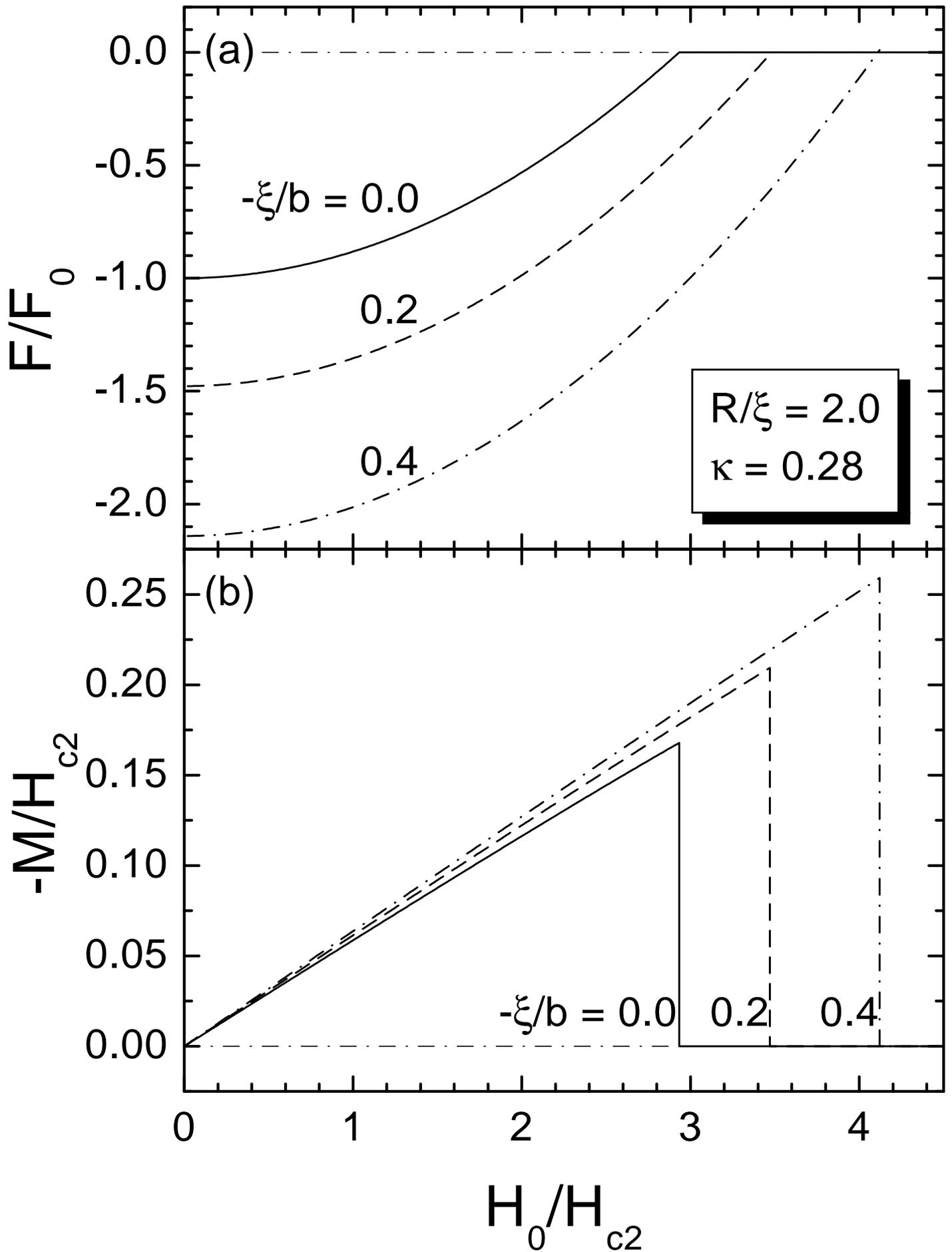
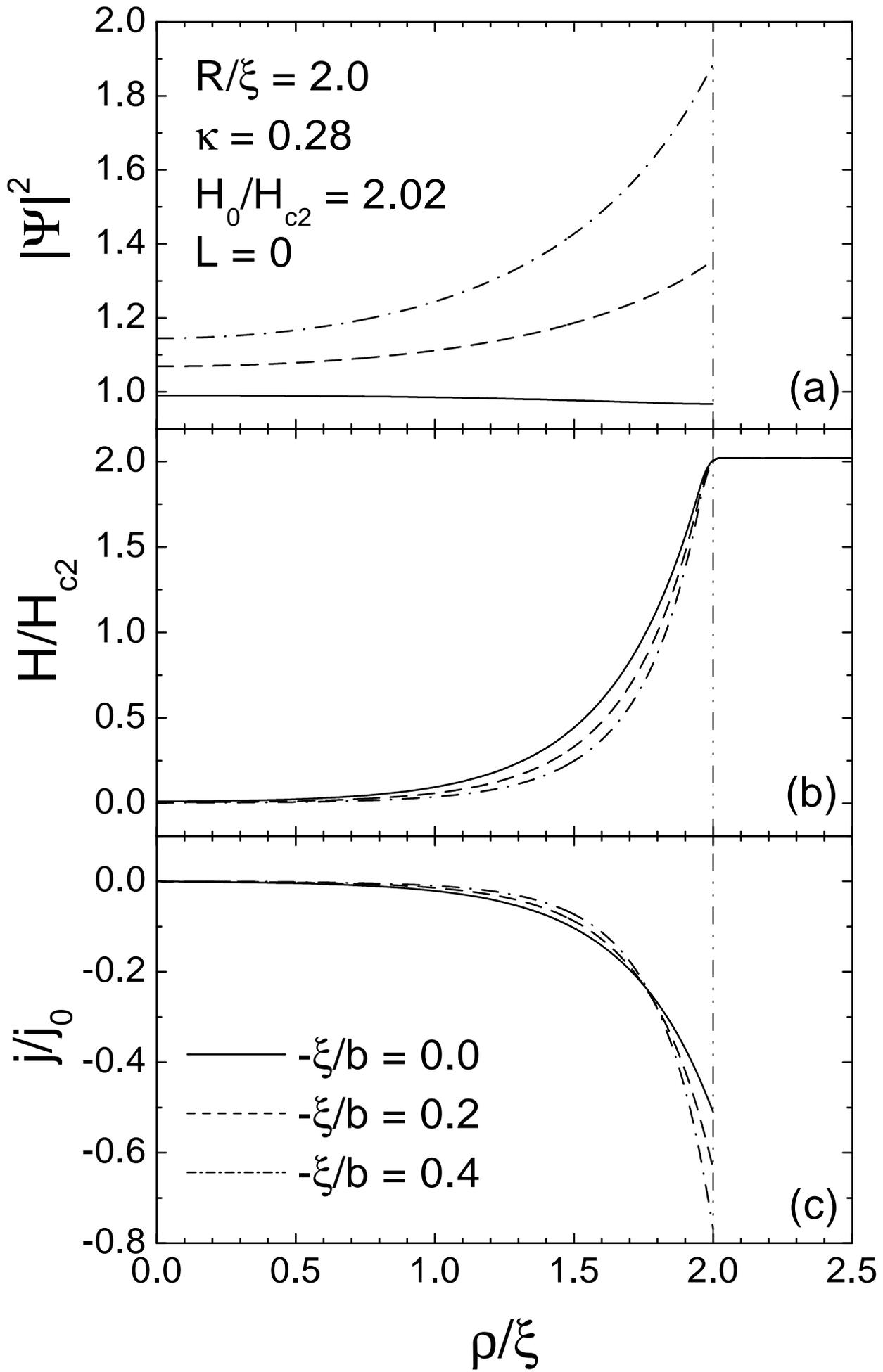
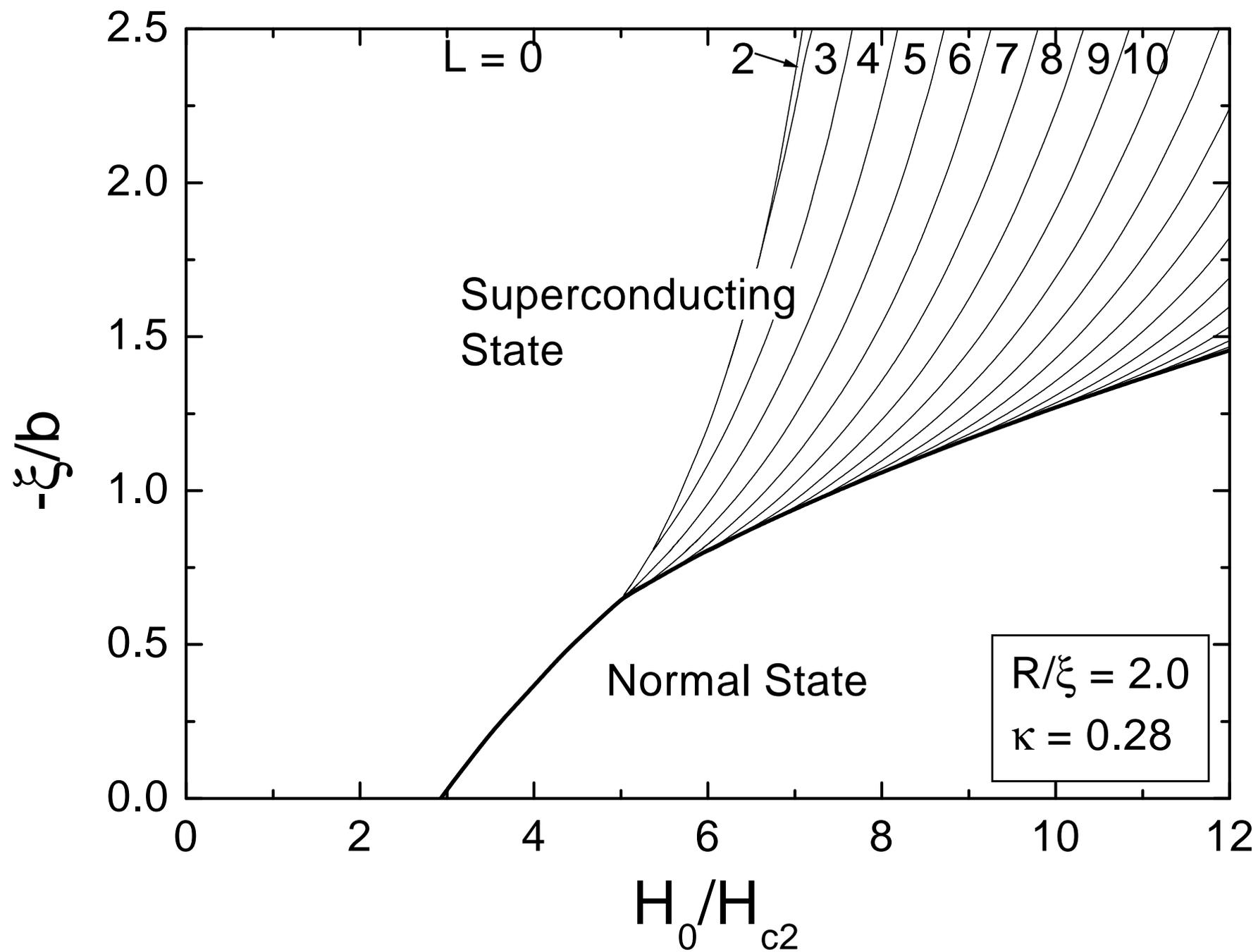
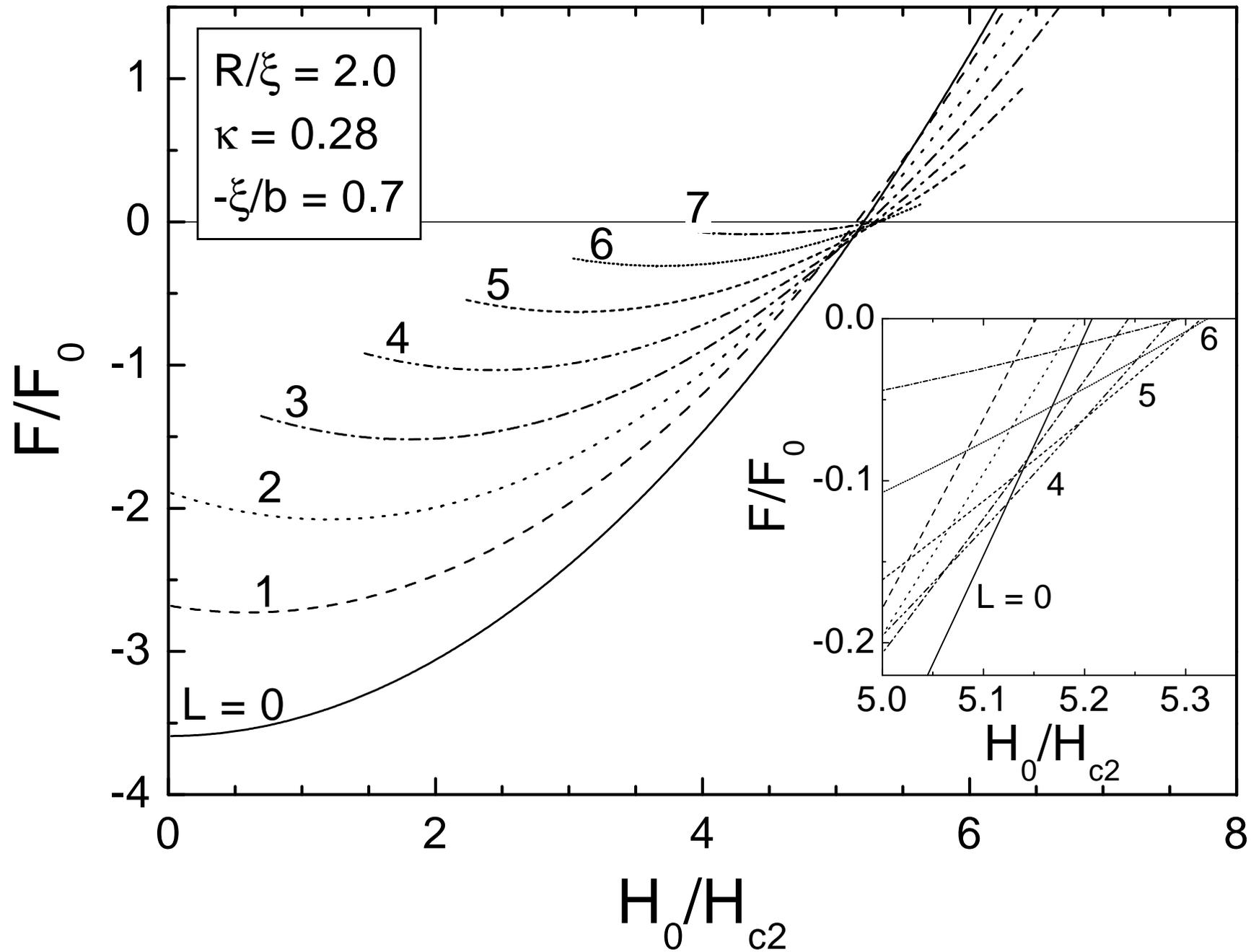
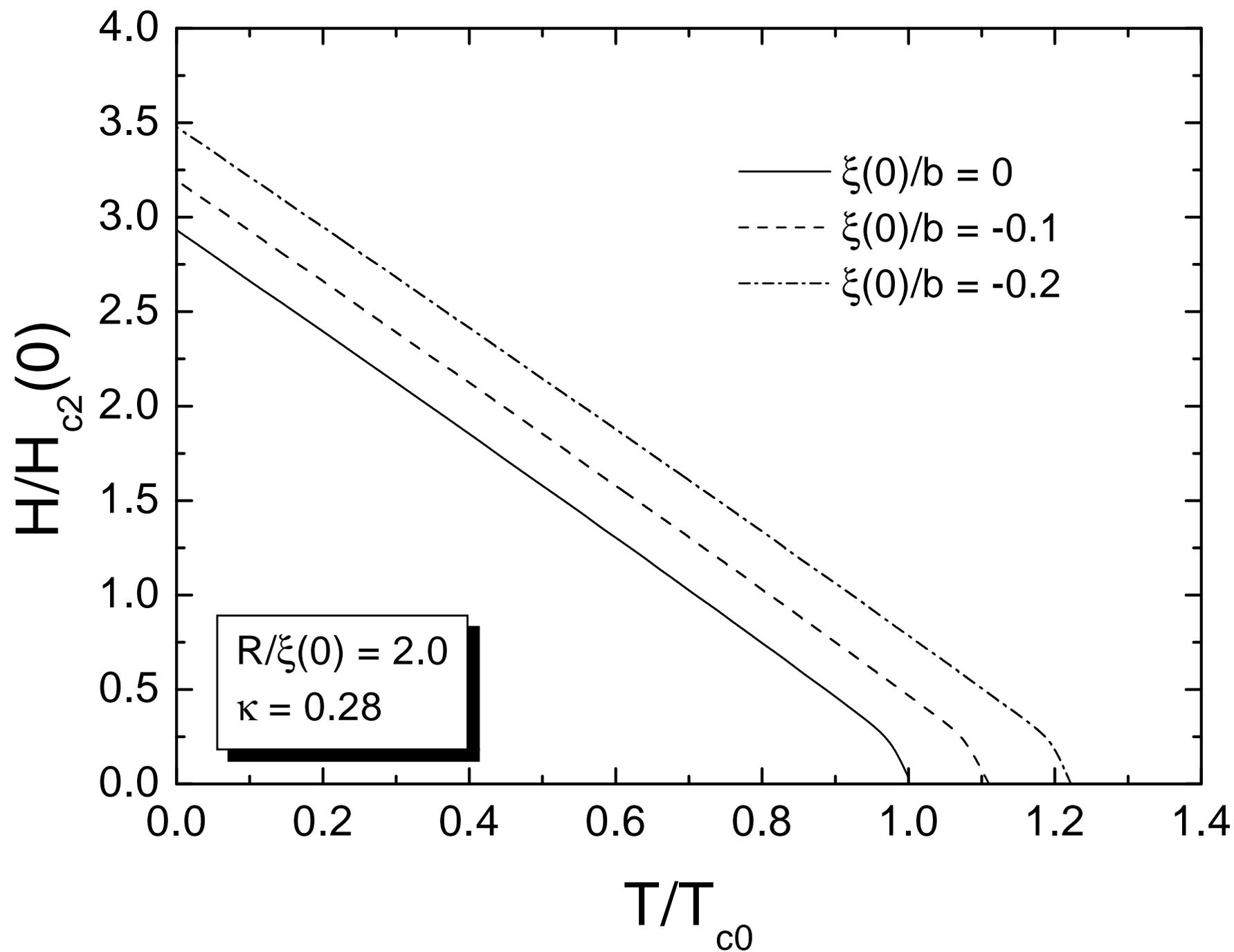
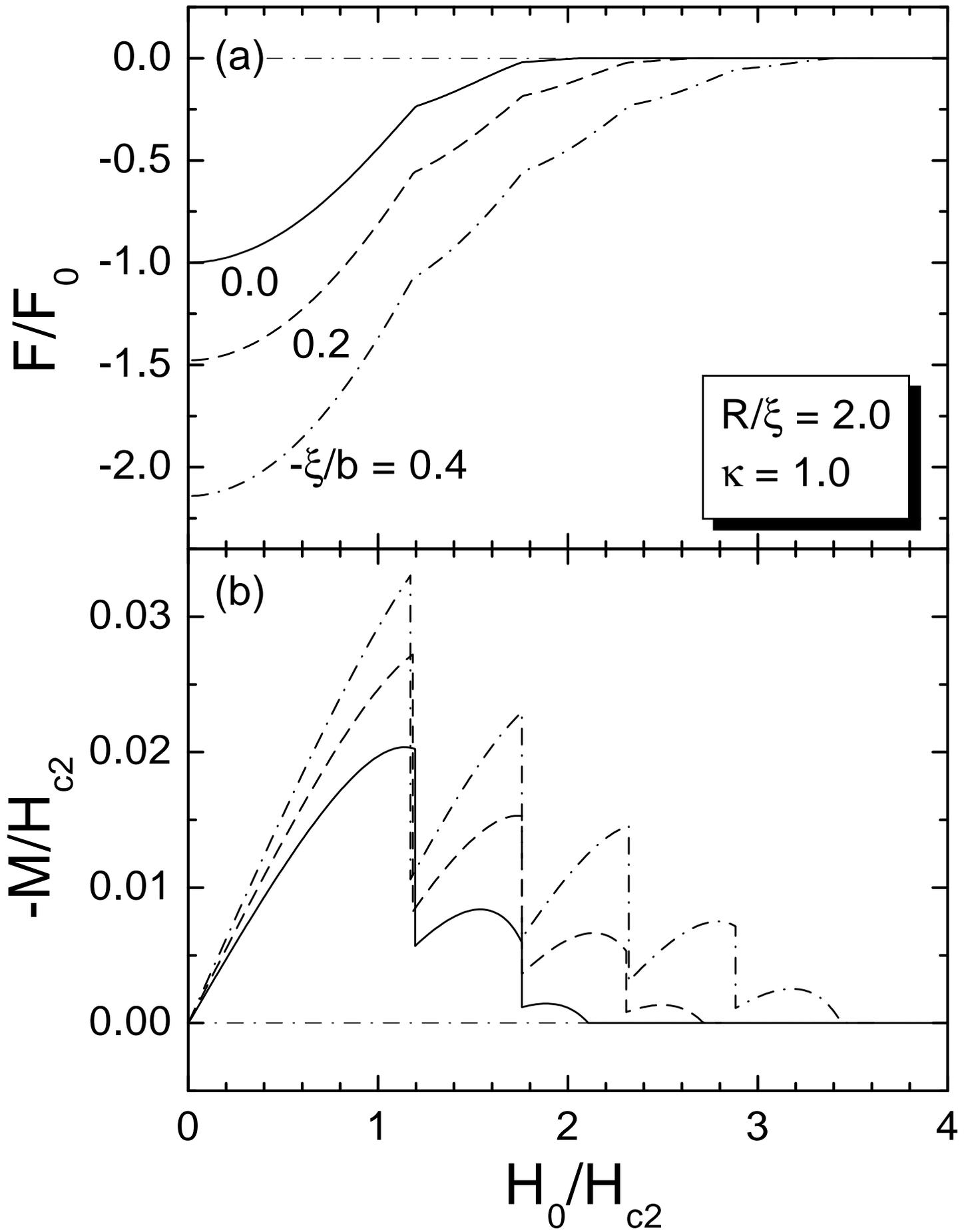
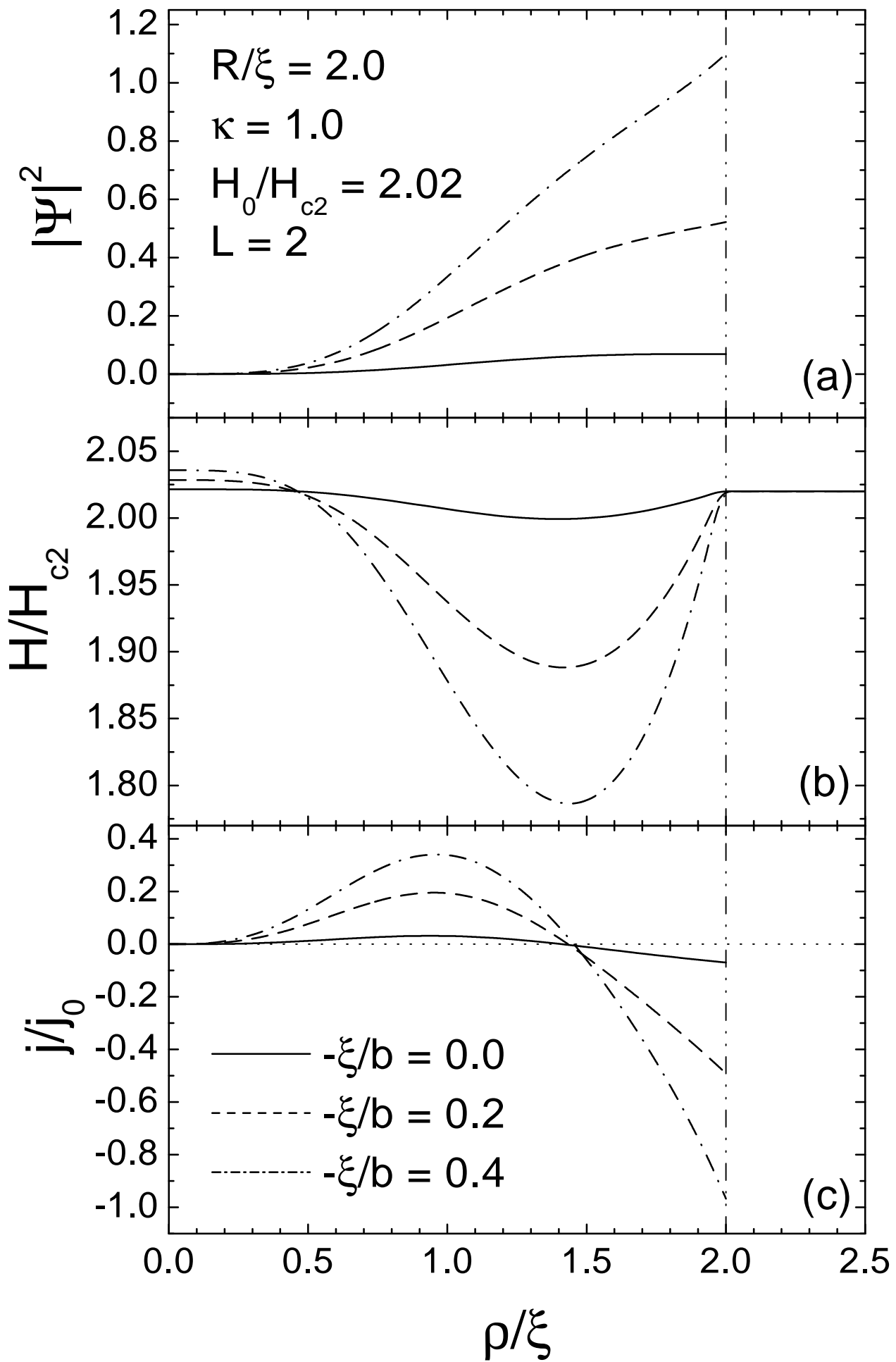
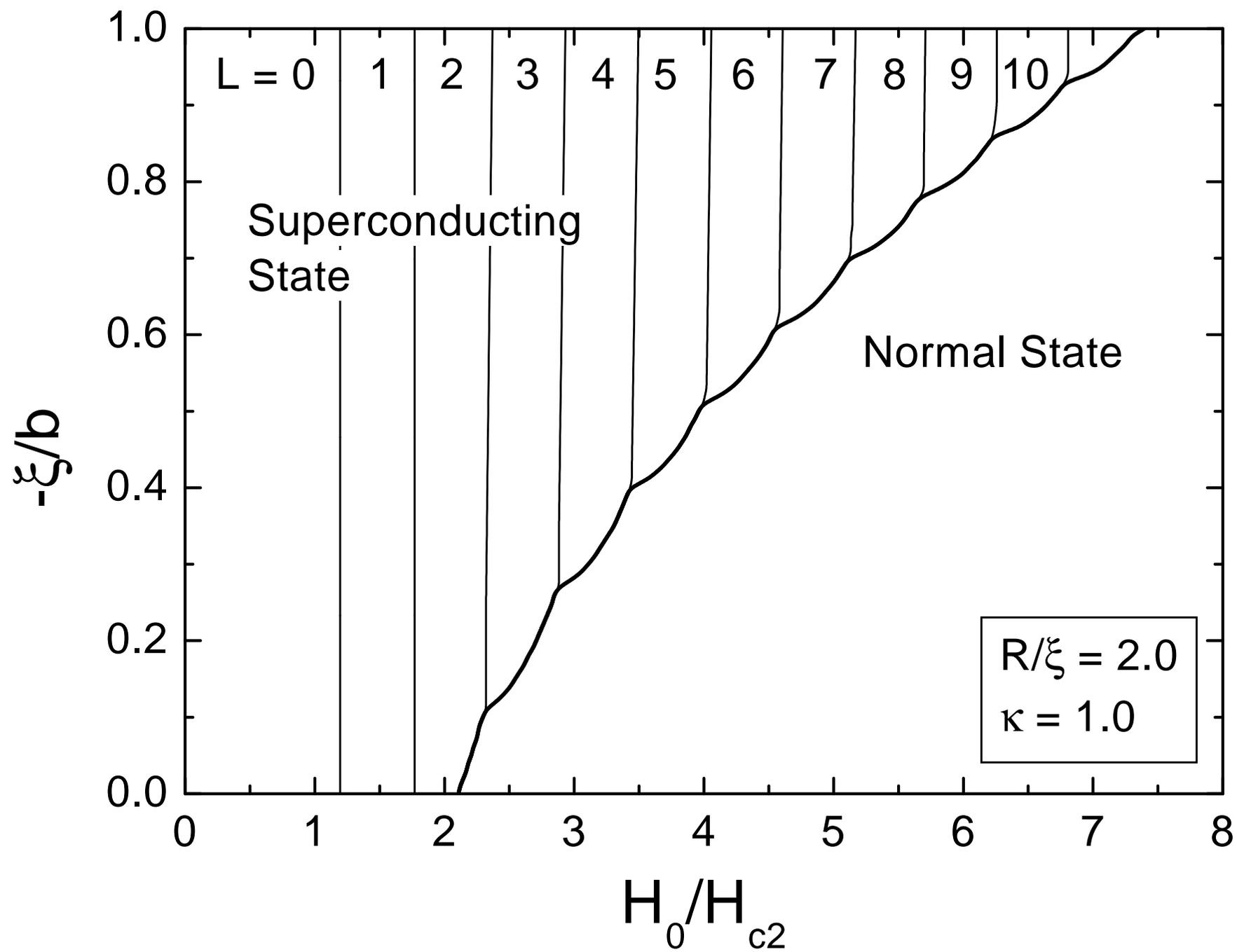
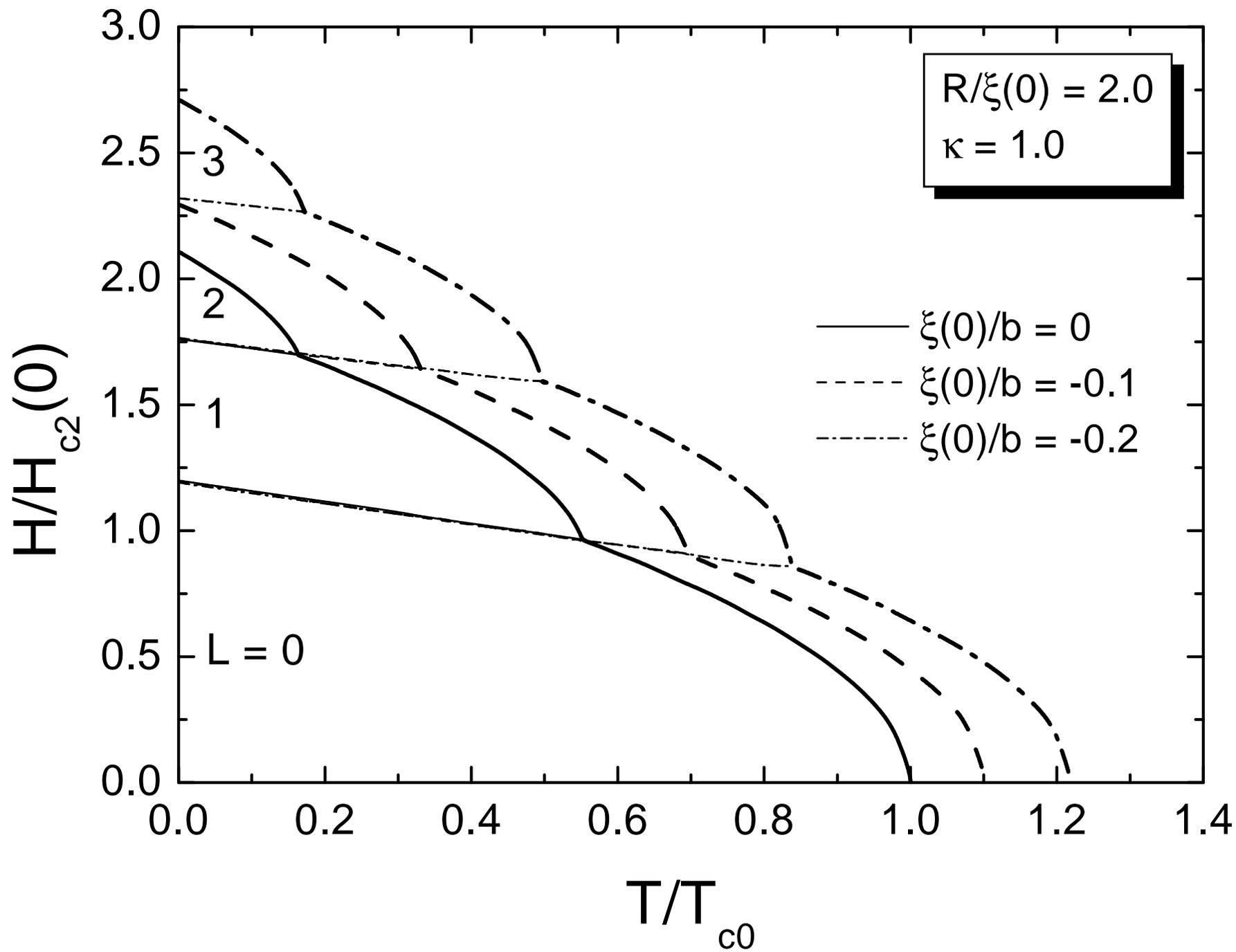
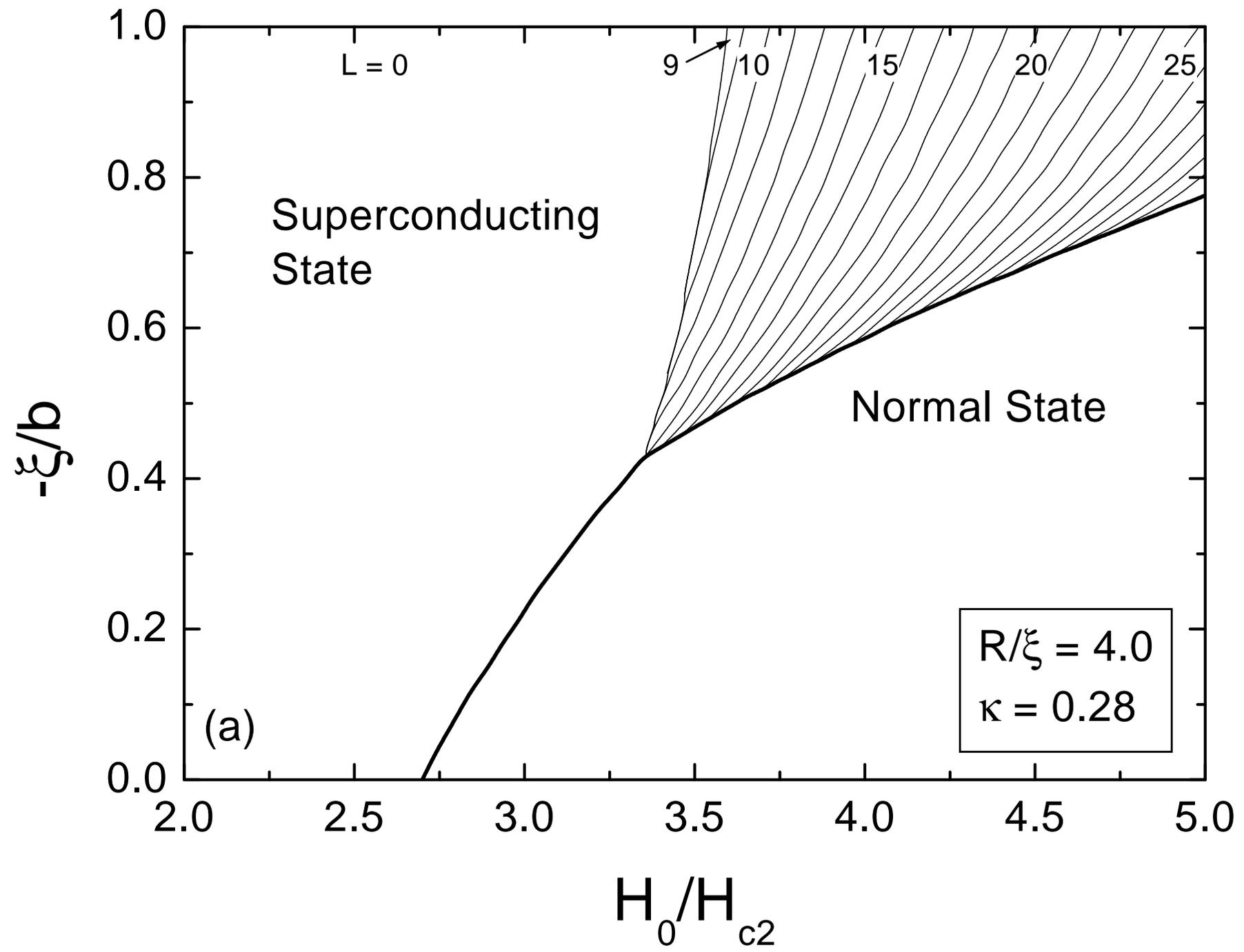
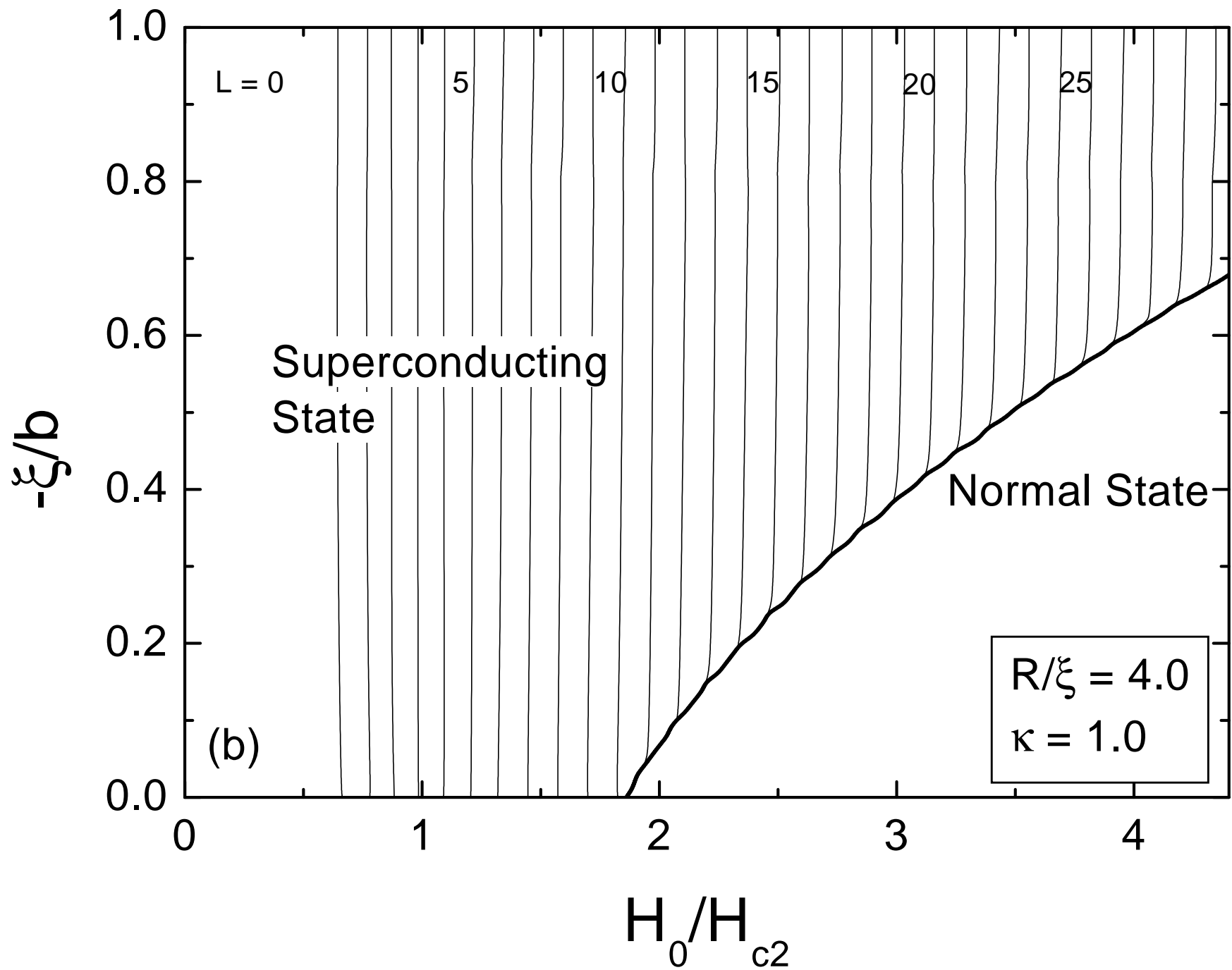
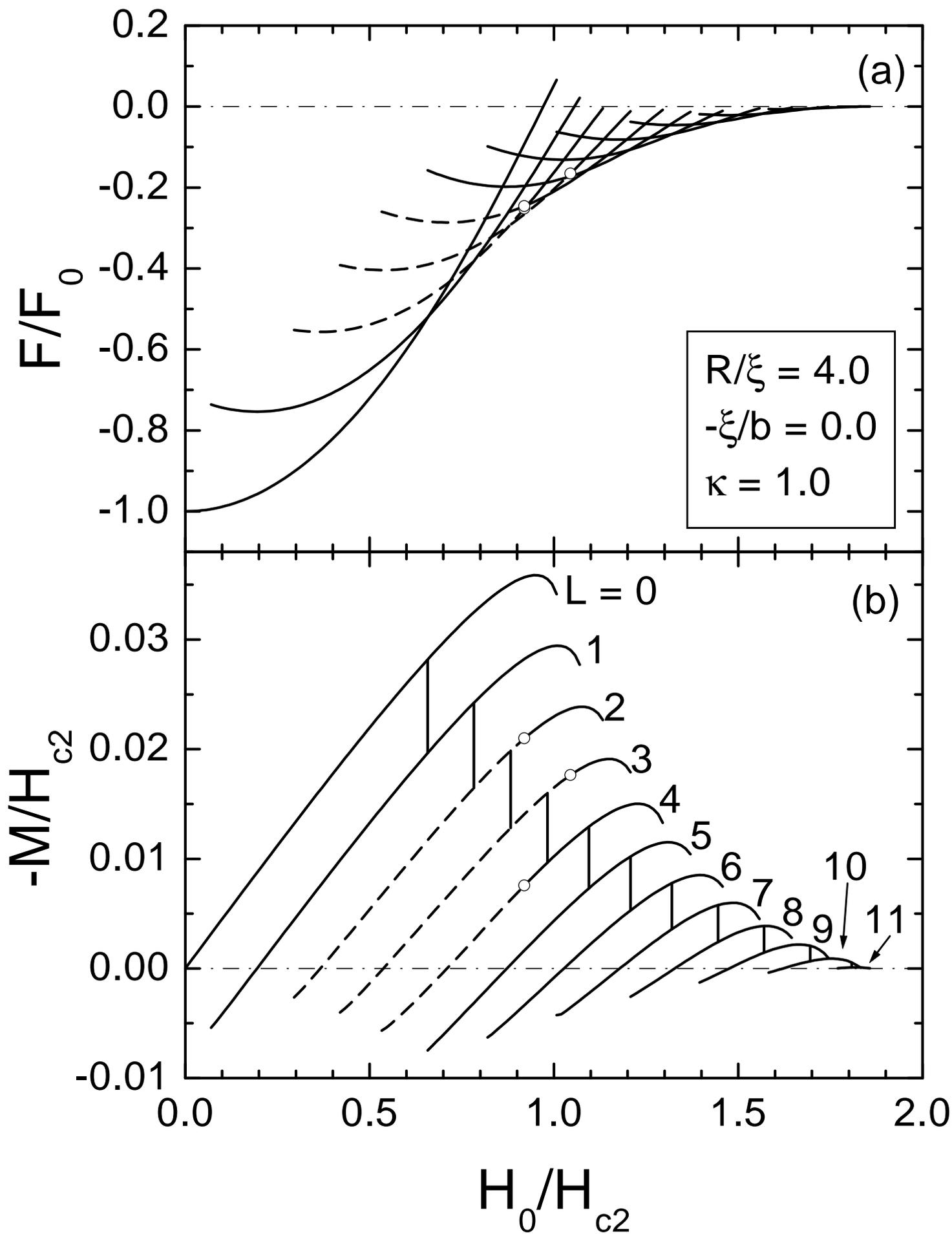
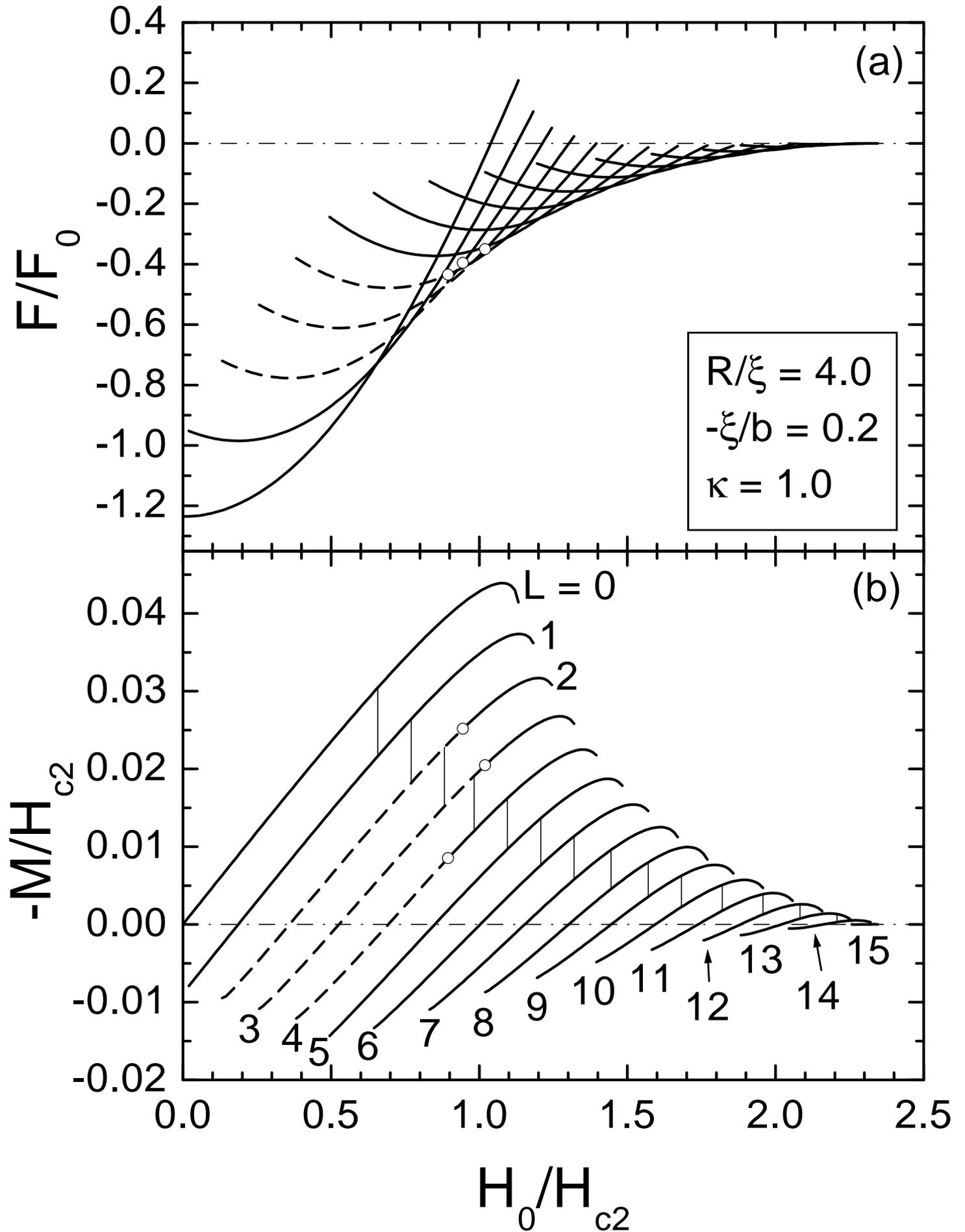
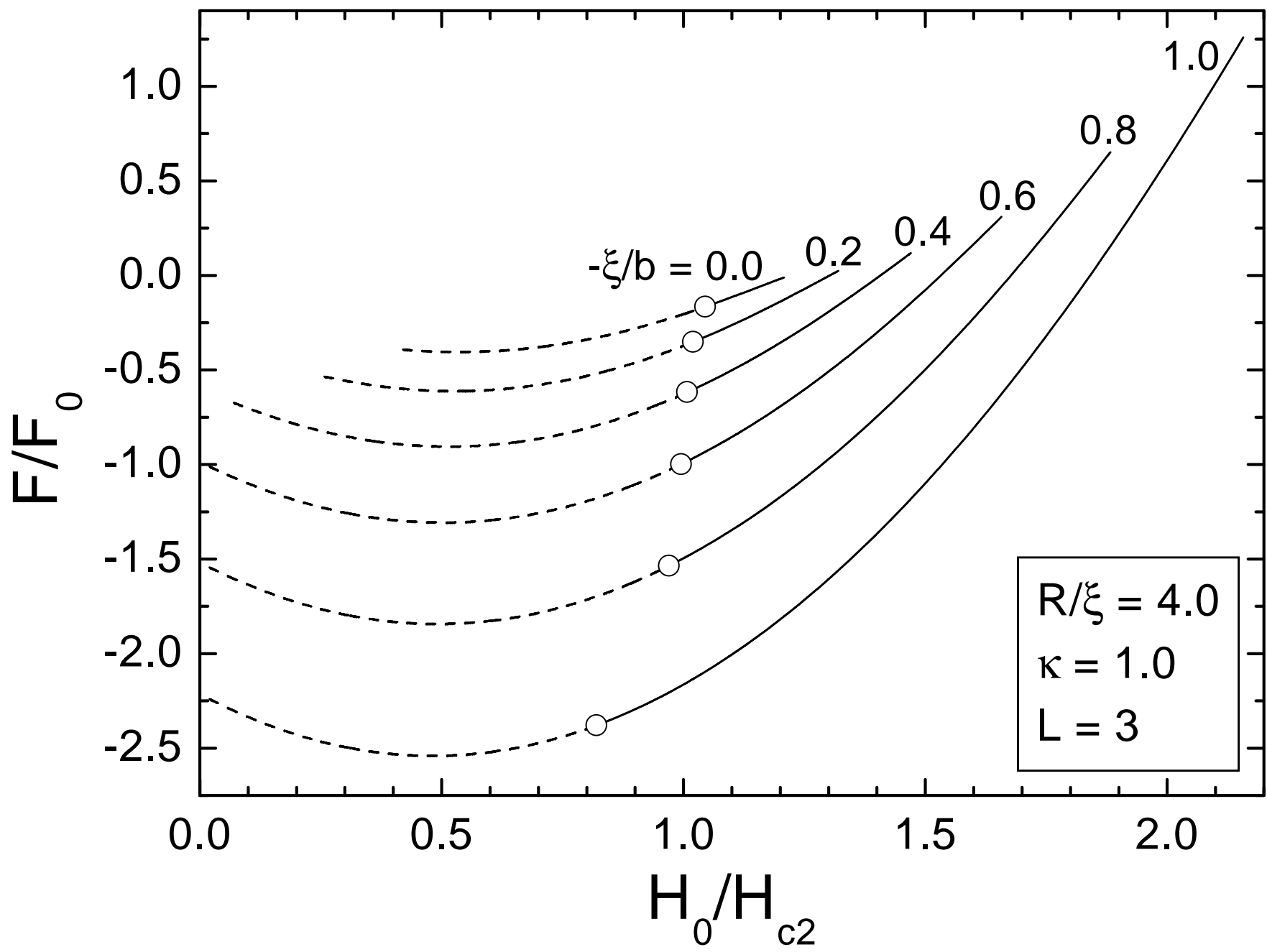
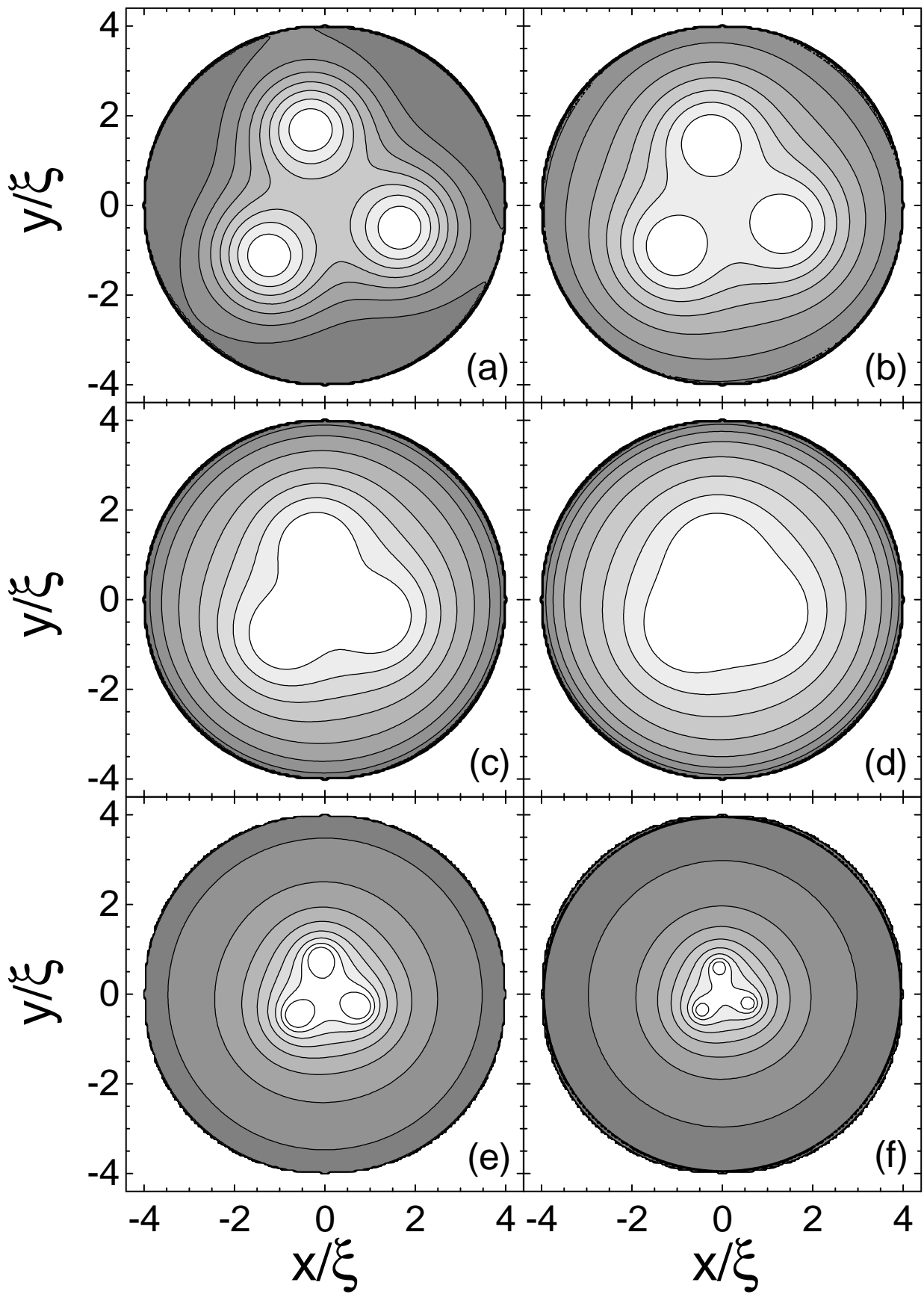
\end{document}